\documentclass[twocolumn,eqsecnum,showpacs,preprintnumbers,amsmath,amssymb,superscriptaddress,nofootinbib,english]{revtex4}

\usepackage{babel}
\usepackage{dcolumn}

\newcommand{\grad}{\mathrm{d}}
\newcommand{\rot}{\mathrm{rot}\,}
\renewcommand{\div}{\mathrm{div}\,}
\newcommand{\laplace}{{\vartriangle}\,}
\newcommand{\eps}{\varepsilon}
\newcommand{\epso}{\varepsilon_{\mathrm{o}}}
\newcommand{\mb}{{\overline{m}}}
\newcommand{\frm}[1]{{\Theta^{(#1)}}}
\newcommand{\NP}[1]{{#1}_{\scriptscriptstyle\mathrm{NP}}}
\newcommand{\EM}{{\scriptscriptstyle\mathrm{EM}}}
\newcommand{\gyr}{{\scriptscriptstyle\mathrm{gyr}}}

\begin{document}

\title{Gyratons on direct-product spacetimes\footnote{%
This is a slightly extended version of the paper published in Phys.~Rev.~D. An additional technical material can be found in sections \ref{ssc:gauge}, \ref{ssc:NPginv} and appendices \ref{apx:NP} and \ref{apx:Green}.}}

\author{Hedvika Kadlecov\'{a}}%
\email[e-mail: ]{hedvika.kadlecova@centrum.cz}
 \affiliation{%
 Institute of Theoretical Physics, Faculty of Mathematics and Physics, Charles University,\\
  V Hole\v{s}ovi\v{c}k\'{a}ch 2, 180 00 Prague 8, Czech Republic
  }%

\author{ Andrei Zelnikov}%
 \email[e-mail: ]{zelnikov@phys.ualberta.ca}
\affiliation{%
Theoretical Physics Institute, Department of Physics, University of Alberta,\\ Edmonton, Alberta, Canada T6G 2J1
}%
\author{Pavel Krtou\v{s}}%
\email[e-mail: ]{Pavel.Krtous@utf.mff.cuni.cz}
 \affiliation{%
 Institute of Theoretical Physics, Faculty of Mathematics and Physics, Charles University,\\
  V Hole\v{s}ovi\v{c}k\'{a}ch 2, 180 00 Prague 8, Czech Republic
}%
\author{Ji\v{r}\'{\i} Podolsk\'{y}}%
\email[e-mail: ]{podolsky@mbox.troja.mff.cuni.cz}
 \affiliation{%
 Institute of Theoretical Physics, Faculty of Mathematics and Physics, Charles University,\\
  V Hole\v{s}ovi\v{c}k\'{a}ch 2, 180 00 Prague 8, Czech Republic
  }%

\date{June 25, 2009}    

\begin{abstract}
We present and analyze exact gyraton and nonexpanding gravitational wave solutions of algebraic type II on backgrounds which are a direct-product of two 2-spaces of constant curvature, or more general type~D spacetimes. This family of electro-vacuum background spacetimes contains the Nariai, anti-Nariai and Pleba\'{n}ski--Hacyan universes, conformally flat Bertotti--Robinson and Minkowski spaces. The gyraton solutions are given in a simple Kundt metric form. They belong to the recently discussed class of spacetimes with constant scalar invariants (CSI) of the curvature tensor. We show that the Einstein equations reduce to a set of linear equations on the transverse 2-space which can be explicitly solved using the Green functions.
\end{abstract}

\maketitle

\section{\label{sec:level1}Introduction}

Recently, there has been a growing interest in investigation of gyraton spacetimes. They represent the gravitational field of a localized source with an intrinsic rotation, moving at the speed of light. Such an idealized ultrarelativistic source, which can be modeled as a pulse of a spinning radiation beam, is accompanied by a sandwich or impulsive gravitational wave. 

In fact, gravitational fields generated by (nonrotating) light pulses and beams were already studied by Tolman \cite{Tol:1934:Oxf:} in 1934, who obtained the corresponding solution in the linear approximation of the Einstein theory. Exact solutions of the Einstein--Maxwell equations for such ``pencils of light'' were found and analyzed by Peres \cite{Peres:1960:PHYSR:} and Bonnor \cite{Bonnor:1969:COMMPH:,Bonnor:1969:INTHP:,Bonnor:1970a:INTHP:}. These solutions belong to a general family of {\it pp\,}-waves \cite{Step:2003:Cam:,GrifPod:2009:Cam:}.

In the impulsive limit (i.e. for an infinitely small cross-section of the beam, and for the delta-type distribution of the light-pulse in time), the simplest of these solutions reduces to the well-known Aichelburg--Sexl metric \cite{Aich-Sexl:1971:} which describes the field of a point-like null particle. It can be obtained by boosting the Schwarzschild metric to the speed of light, with the mass tending to zero so that the total energy is kept finite. More general impulsive waves were subsequently obtained by boosting other black hole spacetimes with rotation, charge and a cosmological constant \cite{FerPen90,LouSan92,HotTan93,KBalNac95,KBalNac96,PodGri97,PodGri98prd} (for recent reviews, see \cite{Podolsky02,BarHog:2003:WorldSci:}).

Gyraton solutions are special sandwich or impulsive waves of the Kundt class (which generalize the {\it pp\,}-waves) such that the corresponding beam of radiation carries not only energy but also an additional angular momentum. Such spacetimes were first considered by Bonnor in \cite{Bonnor:1970b:INTHP:}, who studied the gravitational field created by a spinning null fluid. He called the corresponding particle made out of this continuum a ``spinning nullicon''. In some cases, this may be interpreted as a massless neutrino field \cite{Griffiths:1972a:INTHP:}.

In the exterior vacuum region outside the source, these solutions are locally isometric to standard {\it pp\,}-waves. The interior region contains a nonexpanding null matter which has an intrinsic spin. In general, these solutions are obtained by keeping the nondiagonal terms $g_{ui}$ in the Brinkmann form \cite{Brink:1925:MAAN:} of the {\it pp\,}-wave solution, where $u$ is the null coordinate and $x^i$ are orthogonal spatial coordinates. The corresponding energy-momentum tensor thus also contains an extra nondiagonal term $T_{ui}=j_{i}$. In four dimensions, the terms $g_{ui}$ can be set to zero {\it locally}, using a suitable gauge transformation. However, they can not be {\it globally} removed because the gauge invariant contour integral $\oint g_{ui}(u,x^{j})\,\grad x^{i}$ around the position of the gyraton is proportional to the nonzero angular momentum density $j_{i}$, which is nonvanishing.

Similar gyratons in a higher dimensional flat space were investigated (in the linear approximation) by Frolov and Fursaev \cite{Fro-Fur:2005:PHYSR4:}. Such gyratons represent a pulse of circularly polarized radiation or a modulated beam of ultrarelativistic particles with spin or other sources, which have finite energy~$E$ and finite total angular momentum~$J$. The gyraton itself is characterized by two arbitrary profile functions of $u$ which determine the energy density and angular momentum. The authors investigated the limit in which the source becomes infinitesimally small (with a negligible radius of the cross-section) and the profile functions are independent. They also studied the geodesic motion of test particles in the field of gyraton and demonstrated that, when the gyraton passes through the center of the ring of test particles, the particles start to rotate. In fact, the gyraton's angular momentum effectively creates a force which is similar to the usual centrifugal repulsive force, while the gyraton energy produces the attractive ``Newtonian'' force.

Frolov, Israel, and Zelnikov \cite{Fro-Is-Zel:2005:PHYSR4:} further investigated the exact gyraton solutions propagating in an asymptotically flat D-dimensional spacetime and proved that the Einstein's equations for gyratons reduce to a set of linear equations in the Euclidean ${(D-2)}$-dimensional space. They also showed that the gyraton metrics belong to a class of vanishing scalar curvature invariants (VSI) spacetimes for which all polynomial scalar invariants, constructed from the curvature and its covariant derivatives, vanish identically \cite{Prav-Prav:2002:CLAQG:}. (For the discussion of spacetimes with nonvanishing but nonpolynomial scalar invariants of curvature, see \cite{Page:2009:}.) Subsequently, charged gyratons in Minkowski space in any dimension were presented in \cite{Fro-Zel:2006:CLAQG:}.

In \cite{Fro-Zel:2005:PHYSR4:},  Frolov and Zelnikov took a cosmological constant into account, and exact solution for gyratons in the asymptotically anti-de~Sitter spacetime were presented. Namely, they obtained Siklos gyratons which generalize the Siklos family of nonexpanding waves \cite{Sik:1985:Cam:} (investigated further in \cite{Pod-rot:1998:CLAQG:}). 

In this case, all polynomial scalar invariants are independent of the arbitrary metric functions which characterize the gyraton and have the same values as the corresponding invariants of pure anti-de~Sitter background. The AdS gyratons \cite{Fro-Zel:2005:PHYSR4:} thus belong to the class of spacetimes with constant scalar invariants (CSI) \cite{Coley-Her-Pel:2006:CLAQG,  Coley-Gib-Her-Pope:2008:CLAQG, Coley-Her-Pel:2008:CLAQG, Coley-Her-Pel:2009:CLAQG, Coley-Her-Pap-Pel:2009:}. In string theory it has been demonstrated that generalized {\it pp\,}-wave spacetimes do not get any quantum and $\alpha'$ corrections and hence are perturbatively exact. One may expect a similar property to be valid also for the gyratons, but more careful analy\-sis is required since even if all of the local counterterms in the effective action are trivial constants for CSI spacetimes, their metric variations can be nontrivial functions. Still, one can try to generalize the property of relatively simple quantum corrections to the case of {\it semiuniversal metrics} \cite{ Coley-Gib-Her-Pope:2008:CLAQG} when the Ricci tensor has a block-diagonal structure.

Let us also mention that string gyratons in supergravity were recently found in \cite{Fro-Li:2006:PHYSR4:}.  Supersymmetric gyraton solutions were also obtained for a minimal gauged theory in five dimensions in \cite{Cald-Kle-Zor:2007:CLAQG:}, where the configuration represents a generalization of the Siklos waves with a nonzero angular momentum in anti-de~Sitter space.

The gravitational field generated by gyratons may be interesting for studies of production of mini black holes in colliders (such as the LHC) or in cosmic ray experiments. The problem of mini black hole formation in high energy particle collisions is an important issue of TeV gravity. The theory of such collisions, developed in \cite{Ear-Gid:2002:PHYSR4:,Yosh:2002:PHYSR4:, Yosh:2003:PHYSR4:, Yosh:2005:PHYSR4:, Yosh:2006:PHYSR4:}, was applied to gyraton models in \cite{Yosh:2007:PHYSR4:}.


The purpose of our contribution is to further extend the family of gyratonic solutions, which are only known in Minkowski or anti-de~Sitter background spaces. In particular, we present a new large class of gyratons of algebraic type~II, propagating in less trivial universes which are a direct product of two 2-spaces of constant curvature. This family of vacuum and electrovacuum background spacetimes contains the Nariai \cite{Nariai:1951:}, anti-Nariai, and Pleba\'{n}ski--Hacyan universes  \cite{Pleb-Hacyan:1979:JMATHP:} of type~D, or conformally flat Bertotti--Robinson \cite{Bertotti:1959:,Robinson:1959:} and Minkowski spaces. These direct-product spacetimes with six isometries (see \cite{Step:2003:Cam:,GrifPod:2009:Cam:} for more details) recently attracted new interest because they can be  recovered as specific extreme limits of various black hole spacetimes in four or more dimensions \cite{GinspargPerry:1983:NUCLB:, BousoHawk:1996:PHYSR4:,Car-Dias-Lemos:2004:PHYSR4:,Dias-Lemos-1:2003:PHYSR4:}.

Impulsive gravitational and pure radiation waves in the (anti-)Nariai, Bertotti--Robinson, and Pleba\'{n}ski--Hacyan universes were presented and analyzed by Ortaggio and Podolsk\'{y} \cite{Ortaggio:2002:PHYSR4:, OrtagPodolsky:2002:CLAQG:}.  They showed, and subsequently analyzed in more detail in \cite{PodoOrtag:2003:CLAQG:}, that these solutions are straightforward impulsive limits of a more general class of Kundt spacetimes of type II with an arbitrary profile function, which can be interpreted as gravitational waves propagating on specific type~D or~O backgrounds, including those which are a direct product of two 2-spaces. In fact, the gyraton spacetimes investigated in this paper are generalizations of such Kundt waves when their ultrarelativistic source is made of a ``spinning matter''.

The paper is organized as follows. In Section \ref{sc:gyreq}, we  present the ansatz for the metric and fields. After a short review of the transverse space geometry we derive field equations and simplify them introducing the potentials. Next, we discuss the gauge freedom and suitable gauge fixings. The overview of the gyraton solutions is summarized in Section. \ref{sc:gyrsol}.

In Section \ref{sc:knownsol}, we give a survey of important special subclasses of our gyraton solution. They include direct-product spacetimes, all type~D vacuum backgrounds, and general Kundt waves on these backgrounds.
In Section \ref{sc:interpret}, we concentrate on the interpretation and description of the gyratons. We discuss geometric properties of the principal null congruence, the Newman--Penrose (NP) quantities with respect to natural tetrads, and properties of the electromagnetic field.

The final Section V describes the Green functions required to solve the field equations.
The main results of the paper are summarized in concluding Section \ref{sc:conclusion}. 

Some technical results needed to derive the field equations and NP quantities are left to Appendices \ref{apx:AppA} and \ref{apx:NP}. In Appendix \ref{apx:backgroundsTypeD}, we derive all electro-vacuum solutions of type D, and Appendix \ref{apx:Green} discusses further details concerning the Green functions.

\section{Gyratons on direct product spacetimes}\label{sc:gyreq}

\subsection{The ansatz for the metric and matter}\label{ssc:ansatz}

The aim of this paper is to derive and analyze the family of gyraton solutions describing a gyratonic matter which propagates, together with a related gravitational wave, through a direct-product spacetime filled with a ``uniform'' electromagnetic field.

We assume that such spacetimes belong to the Kundt class. It is characterized by a geometrical property that it admits a nonexpanding, nontwisting, and shear-free null congruence \cite{Step:2003:Cam:,GrifPod:2009:Cam:}. This congruence represents the null direction of propagation of the gyraton and of the accompanying gravitational wave.

In terms of canonical (real) coordinates $\{r,u,x,y\}$, such a metric reads
\begin{equation}\begin{split}\label{m1}
ds^2=&\frac{1}{P^2}\bigl(\grad x^2+ \grad y^2\bigr)-2\,\grad u\,\grad r-2H\,\grad u^2\\
&+2a_{x}\,\grad x\,\grad u+2a_{y}\,\grad y\,\grad u\;,
\end{split}\end{equation}
where ${H(r,u,x,y)}$ can depend on all coordinates, but the functions ${a_x(u,x,y)}$, ${a_y(u,x,y)}$, and ${P(u,x,y)}$ are \mbox{${r}$-independent}.
The restriction ${\partial_r P=0}$ follows from our assumption of vanishing expansion of the Kundt geometry, while the condition ${\partial_r a_i=0}$, where ${i=x,y}$, is necessary here to obtain a gyraton which propagates on a direct-product spacetime background. In fact, this condition is a consequence of the Maxwell equations in the case when the electromagnetic field is present. In the absence of the electromagnetic field, the vacuum Einstein equations admit that functions ${a_i}$ can be linear in~${r}$. However, geometrical properties of such solutions are substantially different from those of the direct-product spacetimes. Therefore, in the following we will always assume that
\begin{equation}\label{aiassumption}
\partial_r a_x=0\;,\quad \partial_r a_y=0\;.
\end{equation}
This assumption thus implies that such solutions belong to the special subclass of Kundt solutions (see section \ref{ssc:prop} for more details).

The metric should satisfy the Einstein equations with a stress-energy tensor generated by the electromagnetic field and the gyraton:
\begin{equation}\label{EinsteinEq}
G_{\mu\nu}+\Lambda g_{\mu\nu}=\varkappa \bigl( T^\EM_{\mu\nu}+T^{\gyr}_{\mu\nu}\bigr)\;.
\end{equation}
Here, $\Lambda$  and  $\varkappa=8\pi G$ are the cosmological and gravitational constants, respectively.

The spacetime can be filled with the background electromagnetic field, which is modified by a gravitational influence of the gyraton. We assume
\begin{equation}\begin{split}\label{EMF}
F&=E\,\grad r\wedge\grad u + B\, \frac{1}{P^2}\grad x\wedge\grad y\\
 &\qquad\qquad+\sigma_{x}\,\grad u\wedge\grad x+\sigma_{y}\,\grad u\wedge\grad y\;,
\end{split}\end{equation}
where $E$ and $B$ are constants, so that the corresponding stress-energy tensor ${T^\EM_{\mu\nu}}$ has the form \eqref{EMT}. This ansatz for the Maxwell tensor has been inspired by the electromagnetic field known in the Bertotti--Robinson \cite{Bertotti:1959:,Robinson:1959:} and Pleba\'{n}ski--Hacyan spacetimes \cite{Pleb-Hacyan:1979:JMATHP:}, to which we have added new terms proportional to functions ${\sigma_i(r,u,x,y)}$. In fact, terms with such a structure are generated if we demand a gauge symmetry of the electromagnetic field under gauge transformation discussed in \ref{ssc:gauge}.

Finally, we must characterize the gyratonic matter by specifying the structure of its stress-energy tensor. It is a generalization of a standard null fluid such that we additionally allow terms corresponding to `internal spatial rotation' of the gyraton source,\footnote{%
In all tensorial expressions for the metric and other symmetric tensors, we understand by, for example, ${\grad u \,\grad x}$ the \textit{symmetric} tensor product ${\frac12(\grad u \otimes\grad x + \grad x \otimes\grad u)}$.}
\begin{equation}\label{m7}
\varkappa\, T^{\gyr}=j_{u}\,\grad u^2+2j_x\,\grad u\,\grad x+2j_y\,\grad u\,\grad y\;.
\end{equation}
We admit a general coordinate dependence of the source functions ${j_u(r,u,x,y)}$ and ${j_i(r,u,x,y)}$. However, it will be shown below that the field equations enforce a rather trivial ${r}$-dependence of these functions. Let us note that previous papers on gyratons, namely \cite{Fro-Is-Zel:2005:PHYSR4:}, \cite{Fro-Zel:2005:PHYSR4:}, \cite{Fro-Zel:2006:CLAQG:}, assumed that the gyraton source is ${r}$-independent.

The gyraton source is thus described only on a phenomenological level, by its stress-energy tensor \eqref{m7}. We do not discuss a possible internal structure of the gyratonic matter, and we do not specify its own field equations. The gyraton stress-energy tensor is assumed to be given, and our aim here is to determine its influence on the metric and the electromagnetic field. However, we have to consider that the gyraton stress-energy tensor is locally conserved. It means that the functions ${j_u}$ and ${j_i}$ must satisfy the constraint given by
\begin{equation}\label{gyrenergycons}
  T^{\gyr}_{\;\,\mu\nu}{}^{\>;\nu}=0\;.
\end{equation}
Of course, if we had considered a specific internal structure of the gyratonic matter, the local energy-momentum conservation would have been a consequence of field equations for the gyraton. Without that, we have to require \eqref{gyrenergycons} explicitly.

To summarize, the fields are characterized by functions ${P}$, ${H}$, ${a_i}$, and ${\sigma_i}$ which must be determined by the field equations, provided the gyraton sources ${j_u}$ and ${j_i}$ and the constants ${E}$ and ${B}$ of the background electromagnetic field are prescribed.

As we will discuss in \ref{ssc:backgrounds}, pure background solutions are obtained when both  gyratons and the gravitational waves are absent, namely for ${T^{\gyr}_{\mu\nu}=0}$, ${a_i=0}$ and ${H \propto r^2}$. For the Minkowski and \mbox{(anti-)Nariai} backgrounds, $T^\EM_{\mu\nu}$ also vanishes, while it is nonzero for the Bertotti--Robinson and Pleba\'{n}ski--Hacyan spacetimes. 

Finally, for later convenience, we introduce a constant~${\rho}$, given by the parameters ${E}$ and ${B}$ of the electromagnetic field,
\begin{equation}\label{rhodef}
\rho=\frac{\varkappa\epso}{2}(E^2+B^2)\;,
\end{equation}
(with ${\varkappa}$ and ${\epso}$ being gravitational and electromagnetic interaction constants,\footnote{%
There are two natural choices of geometrical units: the Gaussian with ${\varkappa=8\pi}$, and ${\epso=1/4\pi}$, and SI-like with ${\varkappa=\epso=1}$.} respectively)
and the constants ${\Lambda_+}$ and ${\Lambda_-}$, defined as
\begin{equation}\label{Lambdadef}
\Lambda_\pm=\Lambda\pm\rho\;.
\end{equation}

\subsection{Geometry of the transverse space}\label{ssc:transsp}

The geometrical structure of the Kundt metric \eqref{m1} identifies the null geodesic congruence generated by $\partial_r$ and parametrized by an affine time $r$, the family of null hypersurfaces ${u=\text{constant}}$, and two-dimensional \textit{transverse spaces} ${r,u=\text{constant}}$. It will be convenient to restrict various equations to these transverse spaces. For example, ${a_i}$ and ${\sigma_i}$ can be understood as components of ${u}$-dependent 1-forms on these two-dimensional spaces. Therefore, we now briefly review some formulas and definitions valid in such two-dimensional transverse geometry.

The transverse space is covered by two  spatial coordinates ${x^i}$, and we use the Latin indices ${i,j,\dots}$ to label the corresponding tensor components. The restriction of the metric \eqref{m1} to the transverse space is
\begin{equation}\label{trmetric}
ds^2_{\perp}=g^{ }_{\!{\perp} ij}\,\grad x^{i}\grad x^{j}=\frac{1}{P^2}\,(\grad x^2+\grad y^2)\;.
\end{equation}
Here we made a useful choice of coordinates ${x^i=\{x,y\}}$ in which ${ds^2_{\perp}}$ has a conformally flat form.\footnote{%
The conformally flat coordinates are not essential, but they simplify some expressions. In a two-dimensional space, a choice of such coordinates is always possible.}

The transverse curvature is fully characterized by the scalar curvature ${R_\perp}$, which in terms of conformally flat coordinates reads (cf.\ the definition \eqref{lapldef} below)
\begin{equation}\label{trsccurv}
\frac12 R_\perp\equiv\laplace\!\log P = P\bigl(P_{,xx}+P_{,yy}\bigr)-\bigl(P_{,x}^2+P_{,y}^2\bigr)\;.
\end{equation}
Inspecting the $ru$ component of the Einstein equations~\eqref{EinsteinEq}, we find that the transverse scalar curvature has to be constant,  
\begin{equation}\label{Peq}
  \frac12 R_\perp = \laplace\!\log P = \Lambda_+\;,
\end{equation}
cf.\ the first lines in equations \eqref{EMT} and \eqref{EinsteinT}, together with \eqref{Lambdadef}. The transverse spaces are thus the constant curvature 2-spaces, all with the \emph{same} curvature. Thanks to this property we can further simplify the choice of the transverse coordinates ${\{x,y\}}$ in such a way that the conformal factor ${P^{-2}}$ in \eqref{trmetric} is ${u}$-independent. Therefore, in the following we may assume
\begin{equation}\label{udepofP}
  \partial_r P= 0\;,\quad \partial_u P=0\;.
\end{equation}

Moreover, using a freedom in the choice of the transverse coordinates, we can also put the conformal factor ${P}$ to a canonical form. There are two standard choices solving \eqref{Peq}, namely,
\begin{equation}\label{P1}
  P=1+\frac14\Lambda_+(x^2+y^2)\;,
\end{equation}
and, for a negative ${\Lambda_+}$,
\begin{equation}\label{P2}
  P=\sqrt{-\Lambda_+}\; x\;.
\end{equation}
However, in the following, we do not need a particular form of ${P}$. It must just satisfy Eq.~\eqref{Peq}.

With the transverse metric \eqref{trmetric} we may associate the Levi-Civita tensor ${\epsilon_{ij}}$ (with ${\epsilon_{xy}=P^{-2}}$) and the covariant derivative denoted by a colon (e.g., ${a_{i:j}}$). We raise and lower the Latin indices using ${g^{ }_{\!{\perp} ij}}$, and we use a shorthand ${a^2\equiv a^i a_i=P^2(a_x^2+a_y^2)}$ for a square of the norm of a \\1-form ${a_i}$. In two dimensions, the Hodge duals of 0-,1- and 2-forms ${\varphi}$, ${a_i}$, and ${f_{ij}}$ read
\begin{equation}
(*\varphi)_{ij} = \varphi\, \epsilon_{ij}\;,\;\;
(*a)_i = a_j \epsilon^j{}_i\;,\;\;
*f = \frac12 f_{ij}\epsilon^{ij} = P^2 f_{xy}\;.
\end{equation}

For convenience, we also introduce an explicit notation for two-dimensional divergence and rotation of a transverse 1-form ${a_i}$,
\begin{align}
  \div a &\equiv a_i{}^{:i} = P^2(a_{x,x}+a_{y,y})\;,\\
  \rot a &\equiv * \grad a = \epsilon^{ij} a_{j,i} = P^2(a_{y,x}-a_{x,y})\;,
\end{align}
and for the Laplace operator of a function ${\psi}$,
\begin{equation}\label{lapldef}
\laplace\psi =  \psi_{:i}{}^{:i} = P^2(\psi_{,xx}+\psi_{,yy})\;.
\end{equation}
Note that the divergence and rotation are related as ${\div a = \rot {*}a}$.

Finally, we will generally assume that the transverse space is topologically simple in the sense that the space of harmonics is trivial. However, sometimes it will be physically relevant to consider also nontrivial solutions of the Laplace equation if we relax the boundary and asymptotical conditions in the noncompact case. For example, a solution of the Laplace equation around a localized source satisfies the homogeneous Laplace equation on the space with the source removed. Such a space is, however, noncompact and the solution is not vanishing on the boundary.

\subsection{The field equations}\label{scc:fequations}

After specifying the ansatz for our fields and reviewing the transverse geometry we can now derive the equations for the gyraton. We have to consider the Einstein equations \eqref{EinsteinEq} together with the Maxwell equations and the condition \eqref{gyrenergycons} for the gyraton source.

We start with the cyclic Maxwell equation. Assuming \eqref{EMF} and \eqref{udepofP} it reads
\begin{equation}\label{MXEC}
  0=\grad F = (\partial_r\sigma_i)\, \grad r\wedge\grad u\wedge\grad x^i -\rot\sigma\, \grad u \wedge \epsilon\;,
\end{equation}
where ${\epsilon=P^{-2}\grad x\wedge\grad y}$. We immediately infer that the 1-form ${\sigma_i}$ is ${r}$-independent, ${\partial_r \sigma_i=0}$, and rotation-free,
\begin{equation}\label{pot1}
  \rot\sigma=0\;.
\end{equation}
The second Maxwell equation ${F_{\mu\nu}{}^{;\nu}=0}$ has only\\ the ${u}$ component nonvanishing,\footnote{%
Here we used that ${a_i}$ and ${\sigma_i}$ are ${r}$-independent. If the condition \eqref{aiassumption} was not assumed before, it would follow from the transverse components of this Maxwell equation.}
which gives
\begin{equation}\label{pot2}
  \div\sigma-E\,\div a + B\,\rot a = 0\;.
\end{equation}
We call \eqref{pot1} and \eqref{pot2} the \emph{potential equations} since they guarantee the existence of potentials which will be discussed in detail in Section \ref{ssc:pots}. For this reason, it is useful to note that these equations imply the conditions 
\begin{equation}\label{potcond}
\begin{gathered}
 \div\bigl[E\,(\sigma{-}E\,a)+B\,{*(\sigma{-}E\,a)}\bigr]=0\;,\\
 \rot\bigl[E\,(\sigma{-}B\,{*a})+B\,{*(\sigma{-}B\,{*a})}\bigr]=0\;,
\end{gathered}
\end{equation}
cf.\ equations~\eqref{dkappadlambda} below.

The Einstein equations can be derived from the Einstein tensor and the electromagnetic stress-energy tensor, which are given in Appendix \ref{apx:AppA}. We have already discussed the $ru$ component which leads to the condition~\eqref{Peq}. The transverse diagonal components $xx$ and $yy$ give
\begin{equation}\label{Einsteinii}
  \partial^2_{r} H  =-\Lambda_-\;.
\end{equation}
We thus obtain the explicit ${r}$ dependence of the metric function ${H}$ as
\begin{equation}\label{Heq}
  H = -\frac12 \Lambda_-\, r^2 + g\,r + h\;.
\end{equation}
where we have introduced ${r}$-independent functions ${g(u,x^j)}$ and ${h(u,x^j)}$.

Finally, the remaining nontrivial components of the Einstein equations are those involving the gyraton source \eqref{m7}. The $ui$ components give an equation related to ${j_i}$, which we call the \emph{first source equation},
\begin{equation}\label{jieq}
\begin{split}
  j_i
  &= \frac12\, f_{ij}{}^{:j}  + g_{,i}-\Lambda_-\, a_i\\
  &\quad+\varkappa\epso\,\bigl[E\,(\sigma_i{-}E\,a_i)+B\,(\sigma_j{-}E\,a_j)\,\epsilon^j{}_i\bigr]\\
  &= \frac12\, f_{ij}{}^{:j}  + g_{,i}-\Lambda_+\, a_i\\
  &\quad+\varkappa\epso\,\bigl[E\,(\sigma_i{-}B\,{a_j}\epsilon^j{}_i)+B\,(\sigma_j{-}B\,a_k\epsilon^k{}_j)\epsilon^j{}_i\bigr]
  \;,
\end{split}\raisetag{44pt}
\end{equation}
where we have introduced the external derivative ${f_{ij}}$ of the 1-form ${a_i}$ as
\begin{equation}\label{fdef}
  f_{ij} = a_{j,i}-a_{i,j} = (*\,\rot a)_{ij} \;.
\end{equation}
For convenience, we have written the equation \eqref{jieq} in two equivalent forms. In the square brackets, they explicitly contain the terms which were already encountered in the equation \eqref{potcond}. We can thus easily split the first source equation into divergence and rotation parts:
\begin{align}
  -\div j &= -\laplace g + \Lambda_-\,\div a\;, \label{divjeq}\\
  -\rot j &= \frac12\laplace b + \Lambda_+\, b \;, \label{rotjeq}
\end{align}
where the function ${b(u,x^j)}$ is the Hodge dual of ${f_{ij}}$,
\begin{equation}\label{bdef}
    b \equiv * f = \rot a\;.
\end{equation}
Equations \eqref{divjeq} and \eqref{rotjeq} carry essentially the same information
as the original source equation \eqref{jieq}.\footnote{%
They are equivalent to \eqref{jieq} if we ignore the possibility of harmonic 1-forms which can exist in topologically nontrivial spaces.}

Next, we examine the condition \eqref{gyrenergycons} for the gyraton source. It gives
\begin{equation}\label{gyrenergycons2}
  -(\partial_r j_i)\,\grad x^i + \bigl(-\partial_r j_u+\div j +a^i \partial_r j_i\bigr)\,\grad u =0\;,
\end{equation}
so that the source functions ${j_i}$ must be ${r}$-independent and ${j_u}$ has to have the structure
\begin{equation}\label{jdecomp}
  j_u = r\,\div j + \iota\;.
\end{equation}
The gyraton source \eqref{m7} is thus fully determined by three \mbox{${r}$-independent} functions ${\iota(u,x^j)}$ and ${j_i(u,x^j)}$.

Finally, from the $uu$-component of the Einstein equation we obtain
\begin{equation}\label{jueq}
\begin{split}
  j_u =\,&\bigl(\laplace g - \Lambda_-\div a)\;r\\
      &+{\!}\laplace h + \frac12  b^2 - \Lambda_- a^2 + 2 a^i g_{,i}\\
      & +\partial_u(\div a) + g\, \div a\\
      &-\varkappa\epso \,(\sigma-Ea)^2\;.
\end{split}
\end{equation}
Comparing the coefficient in front of ${r}$ with \eqref{divjeq}, we find that it consistently reproduces the structure \eqref{jdecomp}. The nontrivial ${r}$-independent part of \eqref{jueq} gives the \emph{second source equation} which can be understood as the equation for the metric function ${h}$,
\begin{equation}\label{heq}
\begin{split}
  \laplace h &=
      \iota \, - \frac12 b^2 + \Lambda_- a^2 - 2 a^i g_{,i} \\
      & +\varkappa\epso(\sigma-Ea)^2
      -\partial_u(\div a)- g\, \div a\;.
\end{split}
\end{equation}

\subsection{Potentials}\label{ssc:pots}

We have thus found that the Maxwell and Einstein equations reduce to two potential equations \eqref{pot1}, \eqref{pot2}, and two source equations \eqref{jieq}, \eqref{jueq}. These equations can further be considerably simplified by introducing potentials for the \mbox{1-forms} ${\sigma_i}$, ${a_i}$ and for the source ${j_i}$.

Indeed, the first potential equation \eqref{pot1} gives immediately that ${\sigma_i}$ has a potential ${\varphi(u,x^j)}$ such that
\begin{equation}\label{phipot}
  \sigma_i = \varphi_{,i}\;.
\end{equation}
Using the Hodge decomposition we can express the 1-form ${a_i}$ using two scalar potentials ${\kappa(u,x^j)}$ and ${\lambda(u,x^j)}$:
\begin{equation}\label{klpotdef}
    a_i = \kappa_{,i}+\epsilon_i{}^j\,\lambda_{,j}\;.
\end{equation}
These potentials control the divergence and the rotation of~${a_i}$ via
\begin{equation}\label{divrota}
    \div a = \laplace\kappa\;,\quad \rot a = -\laplace\lambda   \;.
\end{equation}

Equation \eqref{pot2} imposes a constraint among these three potentials ${\varphi}$, ${\kappa}$, and ${\lambda}$:
\begin{equation}\label{lapphi}
    \laplace\varphi \,=\, \laplace(E\kappa+B\lambda)\;.
\end{equation}
If the transverse space is compact (or if it is noncompact but sufficiently strong asymptotic conditions are imposed) the solution of the Laplace equation is trivial and we immediately obtain
\begin{equation}\label{phisol}
    \varphi = E\,\kappa + B\, \lambda\;.
\end{equation}

By using this constraint, it is possible to show that the potentials ${\kappa}$ and ${\lambda}$ solve the conditions \eqref{potcond}
\begin{equation}\label{dkappadlambda}
\begin{gathered}
  E\,(\sigma_i{-}E a_i)+B\,(\sigma_j{-}E a_j)\,\epsilon^j{}_i =
    (E^2+B^2)\, \lambda_{,j}\epsilon^j{}_i\;,\\
  E\,(\sigma_i{-}B{*a}_i)+B\,(\sigma_j{-}B{*a}_j)\,\epsilon^j{}_i =
    (E^2+B^2)\, \kappa_{,i}\;.
\end{gathered}
\end{equation}

In terms of the potentials, the first source equation \eqref{jieq} can be written as
\begin{equation}\label{jieqpot}
  j_i = \frac12\, f_{ij}{}^{:j} + \Lambda_+\,\lambda_{,j}\, \epsilon^j{}_i -\Lambda_-\,\kappa_{,i}  + g_{,i}\;.
\end{equation}
Its rotation part is the equation \eqref{rotjeq} for ${b}$, the solution of which can be used as a source for the equation for the potential ${\lambda}$,
\begin{equation}
  \laplace\lambda = -b\;.\label{lambdab}
\end{equation}
The divergence part of \eqref{jieqpot} can be written as a relation between the functions ${g}$, ${\kappa}$ and source ${\div j}$,
\begin{equation}\label{kappageq}
  \laplace\!\bigl(g-\Lambda_-\kappa\bigr) = \div j\;.
\end{equation}

The problem further simplifies if we introduce scalar potentials ${p(u,x^j)}$ and ${q(u,x^j)}$ for the gyraton source ${j_i}$,
\begin{equation}\label{sourcepot}
    j_i = p_{,i}+\epsilon_i{}^j\,q_{,j}\;,
\end{equation}
so that
\begin{equation}\label{sourcepotB}
    \rot j = -\laplace q\;,\quad\div j = \laplace p\;.
\end{equation}

Substituting this to the field equation \eqref{jieqpot} and splitting it into the gradient part and the rotation part (i.e., using the Hodge decomposition), we obtain
\begin{equation}\label{kappagp}
    g-\Lambda_-\kappa = p\;,
\end{equation}
and
\begin{equation}\label{lambdaeq}
    \frac12\laplace\lambda+\Lambda_+\lambda=-q \;.
\end{equation}
Let us note that all of the potentials are defined up to an additive constant (which, however, can be ${u}$-dependent). In the derivation of \eqref{kappagp} and \eqref{lambdaeq}, we have absorbed the integration constants into this nonuniqueness of potentials. In view of \eqref{lambdab}, function ${b}$ is then given by
\begin{equation}\label{beq}
    b= 2(\Lambda_+\lambda+q)\;.
\end{equation}

We have thus reduced the field equations to simple algebraical relations \eqref{phisol}, \eqref{kappagp} between the potentials, to the Helmholtz--Poisson equation \eqref{lambdaeq} for~${\lambda}$, and the Poisson equation \eqref{heq} for~${h}$. The last one can be also rewritten using the potentials as
\begin{equation}\label{heqpot}
\laplace\hat h = \iota + q \,\laplace\lambda- p \,\laplace\kappa - 2 a^i p_{,i}\;,
\end{equation}
with ${\hat h}$ closely related to ${h}$:
\begin{equation}\label{hath}
\hat h = h +\partial_u\kappa+\frac12\Lambda_-\kappa^2-\frac12\Lambda_+\lambda^2\;.
\end{equation}

\subsection{Gauge transformation and the field equations in suitable gauges}\label{ssc:gauge}

\subsubsection*{Shift of the ${r}$ coordinate}

To find the gyraton solution explicitly, we need to determine the functions ${h}$, ${g}$, ${a_i}$, and ${\sigma_i}$, provided the gyraton sources ${j_i}$ and ${\iota}$ are prescribed. In terms of the potentials ${\kappa}$, ${\lambda}$, and ${\varphi}$, replacing the transverse 1-forms ${a_i}$ and ${\sigma_i}$, we have obtained equation \eqref{lambdaeq} for ${\lambda}$, \eqref{phisol} for ${\varphi}$, and \eqref{heq} for ${h}$. However, we have only one equation \eqref{kappageq} for ${\kappa}$ and ${g}$.

This deficiency of equations corresponds to the fact that our ansatz \eqref{m1}, \eqref{EMF}, \eqref{m7} admits a gauge freedom.
Indeed, the coordinate transformation ${\tilde r\to r = \tilde r-\psi(u,x^j)}$, accompanied by the following redefinition of the metric functions and fields:
\begin{equation}\begin{gathered}\label{gauge}
r=\tilde r-\psi\;,\\
g=\tilde g-\Lambda_-\psi\;,\quad
h=\tilde h-\frac{1}{2}\Lambda_- \psi^2 + \tilde g\,\psi + \partial_u\psi\;,\\
a_i=\tilde a_i-\psi_{,i}\;,\quad
\sigma_i=\tilde \sigma_i-E\,\psi_{,i}\;,\\
\kappa=\tilde \kappa-\psi\;,\quad
\lambda=\tilde\lambda\;,\quad
\varphi=\tilde\varphi-E\,\psi\;,\\
j_i=\tilde j_i\;,\quad
\iota=\tilde\iota+\psi\, \div j\;,
\end{gathered}\end{equation}
leaves the metric, the Maxwell tensor, and the gyraton stress-energy tensor in the same form. Consequently, all of the field equations remain the same. Such a transformation is a pure gauge transformation and we can use it to simplify the solution of the equations.

This gauge transformation has a  geometrical meaning of shifting the origin of the affine parameter ${r}$ of the null congruence ${\partial_r}$.

Inspecting this gauge transformation, we find that the combination ${\,g-\Lambda_-\kappa\,}$ is gauge invariant. This combination enters the field equation \eqref{kappageq}, and only this combination is thus invariantly determined by the sources, namely, it is equal to ${p}$, cf.~\eqref{kappagp}. The particular splitting into ${g}$ and ${\kappa}$ parts is just a question of the gauge choice.

Indeed, it follows from \eqref{gauge} that it is possible to modify one of the functions $g$, ${\kappa}$, or ${\varphi}$ to an arbitrary value or even to cancel it out from all the equations. Moreover, the freedom to choose one of these functions covers the gauge freedom fully. Therefore, we use them to control the gauge freedom: we may fix the gauge by setting ${g}$, ${\kappa}$, or ${\varphi}$ to be an arbitrarily chosen function. Any of these gauge conditions leads to the same family of solutions, only with a different parametrization of the gauge freedom.

\subsubsection*{Gauge fixing of ${g}$}\label{ssc:g}

Let us start with the gauge condition that the function ${g}$ is an arbitrary function. Then the equation \eqref{kappageq} should be understood as the Poisson equation for~${\kappa}$. In terms of the source potentials we have even the explicit solution given by \eqref{kappagp}. The potential ${\lambda}$ is determined by the equation \eqref{lambdaeq}, and ${\varphi}$ by \eqref{phisol}. Finally, substituting these results into the second source equation \eqref{heq} we obtain the Poisson equation for ${h}$.

In fact, the gauge fixing of ${g}$ can be used to eliminate the metric function ${g}$ completely. Setting
\begin{equation}\label{g0}
  g=0\;,
\end{equation}
the field equation for ${\kappa}$ reduces to
\begin{equation}\label{divjg0}
  \laplace\kappa=-\frac1{\Lambda_-}\,\div j\;,
\end{equation}
with the solution
\begin{equation}\label{kappag0}
  \kappa=-\frac1{\Lambda_-}\,p\;.
\end{equation}
The equation for ${h}$ reads
\begin{equation}\label{heqg0}
\begin{split}
\laplace h &=
      \iota - 2\bigl(\Lambda_+\,\lambda+q\bigl)^2
      + \Lambda_+\, \lambda_{,i}\lambda_{,j}g^{ij}
      + 2\, \lambda_{,i}\, p_{,j}\,\epsilon^{ij} \\
      &\qquad\qquad+\frac1{\Lambda_-}\bigl(p_{,i} p_{,j} g^{ij}+\partial_u\laplace p\bigr)\;,
\end{split}
\end{equation}
and the metric function ${H}$ has only trivial quadratic dependence on ${r}$,
\begin{equation}\label{Hg0}
 H(r,u,x^j)=-\frac12 \Lambda_- r^2 + h(u,x^j)\;.
\end{equation}

\subsubsection*{Gauge fixing of ${\kappa}$}\label{ssc:kappa}

Alternatively, we can fix the potential ${\kappa}$, which is equivalent to the prescription of a value of ${\div a}$. Equation \eqref{kappageq} is then the Poisson equation for ${g}$, otherwise the solution of the field equations proceed in the same way as above.

The special choice
\begin{equation}\label{kappa0}
  \kappa=0\;,\quad\text{i.e.}\quad\div a = 0\;,
\end{equation}
implies simple relations for ${g}$:
\begin{gather}
  \laplace g =\div j\;,\label{divjkappa0}\\
  g = p\;,\label{gkappa0}
\end{gather}
and between the 1-forms ${a_i}$ and ${\sigma_i}$ (or their potentials):
\begin{gather}
  \varphi = B\lambda\;,\label{phikappa0}\\
  \sigma_i = B\, {*a}_i\;.\label{sigmakappa0}
\end{gather}
The equation for ${h}$ now takes the form
\begin{equation}\label{hkappa0}
  \laplace h =
      \iota - 2\bigl(\Lambda_+\,\lambda+q\bigl)^2 + \Lambda_+\, \lambda_{,i}\lambda_{,j}g^{ij}
      + 2\, \lambda_{,i}\, p_{,j}\,\epsilon^{ij} \;.
\end{equation}

\subsubsection*{Gauge fixing of ${\varphi}$}\label{ssc:varphi}

The last natural gauge condition is a fixing of the potential ${\varphi}$. In this case, one first finds the potential ${\lambda}$ by solving the equation \eqref{lambdaeq}. The relation \eqref{phisol} then gives the potential ${\kappa}$. Plugging this into \eqref{kappageq}, the equation for ${g}$ is obtained:
\begin{equation}
  \laplace g =\div j + \frac{\Lambda_-}{E}\,\laplace\varphi-\frac{\Lambda_-B}{E}\,\laplace\lambda\;,\label{gphifixed}
\end{equation}
i.e.,
\begin{equation}
  g =p +\frac{\Lambda_-}{E}\,\varphi-\frac{\Lambda_-B}{E}\,\lambda\;.\label{gphifixed2}
\end{equation}
The 1-form ${a_i}$ can be written in terms of ${\sigma_i}$ and ${\lambda}$ as
\begin{equation}\label{aphifixed}
  E\, a_i =\sigma_i-B\,\lambda_{,i}-E\,\lambda_j\,\epsilon^j{}_i\;.
\end{equation}

Particularly, for
\begin{equation}\label{phi0}
  \varphi=0\;,\quad\text{i.e.}\quad\sigma_i = 0\;,
\end{equation}
we obtain
\begin{gather}
  \kappa=-\frac{B}{E}\, \lambda\;,\label{kappaphi0}\\
  g =p -\Lambda_-\frac{B}{E}\,\lambda\;,\label{gphi0}\\
  E\, a_i =-B\,\lambda_{,i}-E\,\lambda_j\,\epsilon^j{}_i\;.
\end{gather}
The choice ${\varphi=0}$  simplifies the Maxwell tensor \eqref{EMF} to
\begin{equation}\label{EMFphi0}
F=E\,\grad r\wedge\grad u + B\, \frac{1}{P^2}\grad x\wedge\grad y\;.
\end{equation}

\subsubsection*{Reparametrization of the  ${u}$ coordinate}

After the above discussion of the gauge freedom corresponding to the transverse-dependent shift of the ${r}$ coordinate, we should also mention the remaining gauge freedom. The metric \eqref{m1}, the electromagnetic field \eqref{EMF}, and the gyraton stress-energy tensor \eqref{m7} keep the same form under a general reparametrization of the ${u}$ coordinate ${\tilde u\to u=f(\tilde u)}$, accompanied by the rescaling ${\tilde r\to r=\tilde r/f'(\tilde u)}$ of  the ${r}$ coordinate. The metric functions and matter fields must be redefined as
\begin{equation}\begin{gathered}\label{urep}
u=f(\tilde u)\;,\quad
r=\frac{\tilde r}{f'(\tilde u)}\;,\\
g=\frac{\tilde g}{f'(\tilde u)}+\frac{f''(\tilde u)}{f'(\tilde u)^2}\;,\quad
h=\frac{\tilde h}{f'(\tilde u)^2}\;,\\
a_i=\frac{\tilde a_i}{f'(\tilde u)}\;,\quad
\sigma_i=\frac{\tilde \sigma_i}{f'(\tilde u)}\;,\\
\kappa=\frac{\tilde \kappa}{f'(\tilde u)}\;,\quad
\lambda=\frac{\tilde\lambda}{f'(\tilde u)}\;,\quad
\varphi=\frac{\tilde\varphi}{f'(\tilde u)}\;,\\
j_u=\frac{\tilde j_u}{f'(\tilde u)^2}\;,\quad
\iota=\frac{\tilde \iota}{f'(\tilde u)^2}\;,\quad
j_i=\frac{\tilde j_i}{f'(\tilde u)}\;.
\end{gathered}\end{equation}
It is worth to emphasize here that this reparametrization is independent of the transverse spatial coordinates. This gauge transformation is thus `global' from the point of view of the transverse space, and it does not influence the field equations (which we formulated as differential equations on the transverse space) in any significant way.

\subsection{Summary of the gyraton solutions}\label{sc:gyrsol}

Let us now summarize the main equations of the gyratons on direct-product backgrounds. These are spacetimes with the metric of the form
\begin{equation}\begin{split}
ds^2=\frac{1}{P^2}(\grad x^2{+}\grad y^2)-2\,\grad u\,\grad r-2H\,\grad u^2+2a_{i}\,\grad x^i \grad u\;,
\end{split}\end{equation}
filled with the electromagnetic field
\begin{equation}
F=E\,\grad r\wedge\grad u + B\, \frac{1}{P^2}\grad x\wedge\grad y+\grad u\wedge\sigma_{i}\,\grad x^i\;,
\end{equation}
and the gyratonic matter
\begin{equation}
\varkappa\, T^{\gyr}=j_{u}\,\grad u^2+2j_i\,\grad x^i\grad u\;.\end{equation}
The metric function ${H(r,u,x^i)}$ is quadratic in ${r}$,
\begin{equation}
  H = -\frac12 \Lambda_-\, r^2 + g\,r + h\;,
\end{equation}
the gyraton energy density ${j_u(r,u,x^i)}$ can be at most linear in ${r}$,
\begin{equation}
  j_u = r\,\div j + \iota\;,
\end{equation}
and the functions ${g(u,x^i)}$, ${h(u,x^i)}$, ${a_j(u,x^i)}$, ${\sigma_j(u,x^i)}$,
${j_j(u,x^i)}$, and ${\iota(u,x^i)}$ are ${r}$-independent. The function ${P(x^i)}$ is ${r}$ and ${u}$-independent and it satisfies the equation
\begin{equation}
  \laplace\!\log P = \Lambda_+\;.
\end{equation}
It can be solved by ${P}$ of the form \eqref{P1} or, for ${\Lambda_+<0}$ by \eqref{P2}.

The transverse 1-forms ${a_i}$, ${\sigma_i}$, and ${j_i}$ can be written in terms of the scalar potentials ${\kappa}$, ${\lambda}$, ${\varphi}$, ${p}$, and ${q}$ as
\begin{gather}
  \sigma_i = \varphi_{,i}\;,\\
  a_i = \kappa_{,i}+\epsilon_i{}^j\,\lambda_{,j}\;,\\
  j_i = p_{,i}+\epsilon_i{}^j\,q_{,j}\;.
\end{gather}
These potentials are unique up to (unphysical) constants on the transverse space, which can always be gauged away.

Finally, the functions ${h}$, ${g}$, ${\kappa}$, ${\lambda}$, and ${\varphi}$ must satisfy the linear field equations
\begin{gather}
  \varphi = E\,\kappa + B\, \lambda\;,\\
  g-\Lambda_-\kappa = p\;,\label{sumgkpeq}\\
  \frac12\laplace\lambda+\Lambda_+\lambda=-q \;,\label{sumlqeq}\\
  \laplace\hat h = \iota + q \,\laplace\lambda- p \,\laplace\kappa - 2 a^i p_{,i}\;,\label{sumheq}
\end{gather}
where
\begin{equation}\label{sumhhatdef}
  \hat h = h +\partial_u\kappa+\frac12\Lambda_-\kappa^2-\frac12\Lambda_+\lambda^2\;.
\end{equation}

\section{Important special subclasses}\label{sc:knownsol}

The large family of solutions of Einstein--Maxwell equations discussed above belongs to the Kundt class \eqref{m1} of nonexpanding, shear-free and twist-free spacetimes \cite{Kundt:1961:ZEPH:, Kundt:1962:PRS:, Step:2003:Cam:}, namely to its subclass characterized by the condition \eqref{aiassumption}. As we have seen in Section \ref{ssc:ansatz}, and will be discuss more in Section \ref{sc:interpret}, the gyratonic matter \eqref{m7} is the ``rotating'' generalization of a null fluid. As special cases, this family of solutions contains some previously known spacetimes from the Kundt family which correspond to electro-vacuum or pure (null) radiation. In this section we will  shortly discuss such important subcases.

\subsection{Direct-product background spacetimes}\label{ssc:backgrounds}

It is natural to start with the simplest case of highly symmetric spacetimes. Considering the metric function ${H}$ of the form \eqref{Heq}, setting ${\,a_{i}=0}$, and ${\,g=0=h}$,
and choosing the expression \eqref{P1} for ${P}$, the metric \eqref{m1} reduces to
\begin{equation}\label{n3}
d s^2=\frac{\grad x^2+ \grad y^2}{[1+\frac{1}{4}\Lambda_{+}\,(x^2+y^2)]^2}-2\grad u\,\grad r+\Lambda_{-}\,r^2\grad u^2\;.
\end{equation}
It describes backgrounds on which the gyratons propagate. By performing the transformation ${r=v(1-\frac12\Lambda_{-}\, u v)^{-1}}$ with ${u=(t-z)/\sqrt 2}$, and ${v=(t+z)/\sqrt 2}$, the metric becomes
\begin{equation}\label{n3tr}
d s^2=\frac{\grad x^2+ \grad y^2}{[1+\frac{1}{4}\Lambda_{+}\,(x^2+y^2)]^2}
+\frac{\grad z^2-\grad t^2}{[1+\frac{1}{4}\Lambda_{-}\,(z^2-t^2)]^2}\;.
\end{equation}
The background spacetimes thus have geometry of a direct product of two 2-spaces of constant curvature $\Lambda_{+}$ and $\Lambda_{-}$, respectively. The first is the space spanned by two spatial coordinates so that it is flat Euclidean space $E^{2}$, {2-sphere} $S^{2}$, or 2-hyperboloid $H^{2}$, according to the sign of the constant $\Lambda_{+}$. The second is the (1+1)-dimensional spacetime spanned by a timelike coordinate and one spatial coordinate. According to the sign of the constant~$\Lambda_{-}$, it is Minkowski 2-space $M_{2}$, de Sitter space $dS_{2}$, or anti-de~Sitter space $AdS_{2}$.

Therefore, there are nine theoretically possible distinct subclasses given by the choice of $\Lambda_{+}$ and $\Lambda_{-}$, but only six of them are physically relevant because the energy density $\rho$ must be non-negative, which eliminates three cases. The most important of such background spacetimes are summarized in Table~\ref{table1}. In addition, there are more general Bertotti--Robinson direct product space-times for which the constants $\Lambda_{+}$ and $\Lambda_{-}$ are independent and nontrivial; i.e., the cosmological constant $\Lambda$ and the energy density ${\rho>0}$ of the electromagnetic field can be chosen arbitrarily.

\begin{table}
\caption{\label{table1}Some of possible background spacetimes which are the direct product of two 2-spaces of constant curvature. Here $\Lambda$ is a cosmological constant, and $\rho$ is a constant energy density of the electromagnetic field.}
\begin{ruledtabular}
\begin{tabular}{cccccccc}
 $\Lambda_{+}$ &  $\Lambda_{-}$ & \text{geometry} & \text{spacetime}  & $\Lambda$ & $\rho$\\
\hline
 0 & 0                  & ${E^{2}\times M_{2}}$  & Minkowski        & $=0$ & $=0$ \\
 $\Lambda$ & $\Lambda$  & ${S^{2}\times dS_{2}}$  & Nariai          & $>0$ & $=0$ \\
 $\Lambda$ & $\Lambda$  & ${H^{2}\times AdS_{2}}$  & anti-Nariai   & $<0$ & $=0$ \\
 $\rho$ & $-\rho$ & ${S^{2}\times AdS_{2}}$  & Bertotti--Robinson   & $=0$ & $>0$ \\
 $2\Lambda$ & 0   & ${S^{2}\times M_{2}}$  & Pleba\'{n}ski--Hacyan   & $>0$ & $=\Lambda$ \\
 0 &  $2\Lambda$  & ${E^{2}\times AdS_{2}}$  & Pleba\'{n}ski--Hacyan & $<0$ & $=|\Lambda|$
\end{tabular}
\end{ruledtabular}
\end{table}

In a natural null tetrad, the only nonvanishing NP Weyl and curvature scalars are (see Section~\ref{ssc:NP})
\begin{equation}\begin{aligned}\label{f18back}
\Psi_{2}&=-\frac{1}{3}\Lambda\,, \qquad R=4\Lambda\,, \qquad \Phi_{11}=\frac{1}{2} \rho\,,\\
\end{aligned}\end{equation}
where ${\Lambda=\frac12(\Lambda_{+}+\Lambda_{-})}$ and ${\rho=\frac12(\Lambda_{+}-\Lambda_{-})}$, together with ${\Phi_{1}=\frac{1}{2} (E+iB)}$. These electro-vacuum solutions are thus of algebraic type~D, unless ${\Lambda=0}$ which applies to a conformally flat Bertotti--Robinson universe and flat Minkowski space. Vacuum direct product spacetimes (with ${\rho=0}$) are Minkowski and (anti-)Nariai spaces. For the two Pleba\'{n}ski--Hacyan spacetimes, one and only one of the 2-spaces is flat. Therefore, ${\Phi_{11}=\tfrac{1}{2}|\Lambda|}$, so that the condition $2\Phi_{11}\pm 3\Psi_{2}=0$ is satisfied.

More details about some of these background spacetimes can be found in the original works \cite{Nariai:1951:,Bertotti:1959:,Robinson:1959:,Pleb-Hacyan:1979:JMATHP:}, reviews \cite{Step:2003:Cam:, GrifPod:2009:Cam:} or, e.g., in \cite{Ortaggio:2002:PHYSR4:,OrtagPodolsky:2002:CLAQG:,Dias-Lemos:2003:PHYSR4:}.

\subsection{Type D background spacetimes}\label{ssc:backgroundsTypeD}

As will be seen in Section \ref{ssc:NP}, the gyraton spacetimes are in general of algebraic type~II. However, they contain a wider subclass of electro-vacuum solutions of type~D, which can also be naturally regarded as possible background geometries.

Type~D electro-vacuum solutions of Einstein's equations are known \cite{Step:2003:Cam:,GrifPod:2009:Cam:,Kinnersley,Carter,Plebanski:1979,Pleb-Hacyan:1979:JMATHP:}. However, their forms are usually different from the  parametrization of the geometry used here. For this reason we will write those type D spacetimes, which belong to our subclass of the Kundt family, explicitly. All such spacetimes are derived in Appendix \ref{apx:backgroundsTypeD}; here we only summarize the results.

Although these spacetimes have the same curvature scalars as in \eqref{f18back}, they are not, in general, direct-product spaces. In particular, they have a lower symmetry than the highly symmetric backgrounds discussed above.

\subsubsection*{The $\Lambda_+=0$ case\\(exceptional Pleba\'{n}ski--Hacyan spacetime)}

As shown in Appendix \ref{apx:backgroundsTypeD}, all type D solutions naturally split into two cases. For ${\Lambda_+=0}$ (i.e., ${\Lambda=-\rho<0}$, ${\Lambda_-=2\Lambda}$) we find a generalization of the exceptional Pleba\'{n}ski--Hacyan type~D electro-vacuum spacetime \cite{Pleb-Hacyan:1979:JMATHP:,OrtagPodolsky:2002:CLAQG:,PodoOrtag:2003:CLAQG:}. The metric reads
\begin{equation}\label{genPlHac}
\begin{split}
d s^2&=
   \grad x^2+ \grad y^2
   +2\bigl(\Lambda\, r^2 - L_{x}\,x - L_{y}\,y\bigr) \grad u^2\\
   &\qquad\qquad\quad
   -2\grad u\grad r
   +2\bigl(a_x \grad x  + a_y\grad y \bigr)\grad u
   \;,
\end{split}
\end{equation}
where ${L_{i}(u)}$ and ${a_i(u)}$, ${i=x,y}$, are arbitrary functions of the coordinate $u$ only (i.e., constants on each transverse space). This corresponds to the metric \eqref{m1} with ${P=1}$, ${g=0}$, ${h}$ linear in ${x,y}$, and ${a_i}$ independent of ${x,y}$.

For ${a_i=0}$ it reduces to the exceptional Pleba\'{n}ski--Hacyan spacetime. It further reduces to the direct product spacetime~\eqref{n3} when also both ${L_{i}}$ vanish. Although the functions ${L_{i}\not=0}$ do not enter the curvature scalars \eqref{f18back}, the geometry of this spacetime is different from that of the direct-product spacetimes (for example, it contains another shear-free but non-geodesic null direction).

Nontrivial coefficients ${a_i}$ can be gauged away using the transformation \eqref{gauge}. However, such a transformation generates a nonvanishing metric function ${g}$ and a quadratic dependence of ${h}$ on ${x,y}$. It thus seems that the case ${a_i\neq0}$ is indeed a nontrivial generalization of the exceptional Pleba\'{n}ski--Hacyan spacetime.

\subsubsection*{The $\Lambda_+\neq0$ case}

In the case when the transverse space has a nonvanishing curvature ${\Lambda_+}$, the metric of  type D electro-vacuum solutions is given by the metric functions
\begin{gather}
  \lambda=\frac{Q}{P}\;,\quad\kappa=0\;,\label{gentypeDkl}\\
  h=\frac12\Lambda_+\lambda^2\;,\quad g=0\;.\label{gentypeDh}
\end{gather}
Here the functions ${P}$ and $Q$ can be written as
\begin{equation}\label{gentypeDPQ}
\begin{gathered}
  P=1+\frac14\Lambda_+(x^2+y^2)\;,\\
  Q=q_0\bigl(1-\frac14\Lambda_+(x^2+y^2)\bigr)+q_x x +q_y y\;,
\end{gathered}
\end{equation}
respectively, where ${q_0(u),\,q_x(u)}$ and $q_y(u)$ are constant on the transverse space.

In the case ${\Lambda_+>0}$, when the transverse space is a sphere, the solution for ${\lambda}$ can also be rewritten as 
\begin{equation}\label{reglambdasph}
\lambda=C\,\bigl[\cos\theta\cos\theta'+\sin\theta\sin\theta'\cos(\phi-\phi')\bigr]\;,
\end{equation}
where ${\theta}$ and ${\phi}$ are standard spherical coordinates, cf.~\eqref{s2}, and ${C}$, ${\theta'}$, and ${\phi'}$ are (possibly ${u}$-dependent) transverse constants equivalent to ${q_0,\,q_x}$  and ${\,q_y}$.

A slightly more general parametrization of these spacetimes can be found in Appendix \ref{apx:backgroundsTypeD}.

\subsection{Kundt waves without gyratons (${j_i=0}$)}\label{ssc:KundtWaves}

\subsubsection*{Kundt waves on direct product spacetimes}

Now we briefly describe more general Kundt spacetimes of the form \eqref{m1} which, however, still do \emph{not} contain a gyratonic matter. In such a case, the source functions $j_i$ vanish, i.e., ${p, q=0}$, and the field equation~\eqref{lambdaeq} reduces to
\begin{equation}\label{vaclambdaeq}
  \laplace \lambda + 2\Lambda_+\, \lambda =0\;.
\end{equation}
Let us first consider the trivial solution ${\lambda=0}$; the general case is discussed below.

Since ${p=0}$, it is possible to use the gauge transformation \eqref{gauge} to eliminate both ${\kappa}$ and ${g}$, cf.~\eqref{kappagp}. Consequently, we obtain ${a_i=0}$ everywhere, and the metric simplifies to
\begin{equation}\label{n3waves}
d s^2=d s^2_{\mathrm{bg}} -2\,h(u,x, y)\,\grad u^2\,,
\end{equation}
where $d s^2_{\mathrm{bg}}$ is the metric of direct product spacetimes (some of which are listed in Table~\ref{table1}) given by \eqref{n3}. For nontrivial profile functions $h$, this class of solutions can be interpreted as specific exact Kundt gravitational waves which propagate in flat, (anti-)Nariai, Bertotti--Robinson, or  Pleba\'{n}ski--Hacyan universes (see \cite{PodoOrtag:2003:CLAQG:} and, for the limit of impulsive waves, \cite{Ortaggio:2002:PHYSR4:,OrtagPodolsky:2002:CLAQG:}).

Indeed, from the corresponding NP scalars \eqref{RicciPhipot}, \eqref{WeylPsipot} (cf. Section~\ref{ssc:NP}), by using \eqref{Delta} it follows that \eqref{f18back} remains unchanged, and, in addition, there is
\begin{equation}\label{f17alter}
\Psi_{4}=(P^2h_{,\zeta})_{,\zeta}\,,\quad\Phi_{22}=P^2h_{,\zeta\bar{\zeta}}\,,
\end{equation}
where ${\zeta=(x+iy)/\sqrt{2}}$. When ${\Psi_{4}\not=0}$, such spacetimes are thus of types~II or~N, and in general, contain a null radiation field characterized by $\Phi_{22}$. In particular, \emph{pure vacuum} gravitational waves of this type (which propagate on a vacuum or electrovacuum background space) are given by the condition ${\Phi_{22}=0}$, so that their profile functions $h$ must be of the form
\begin{equation}\label{vacuumwaves}
h=\mathcal{F}(u,\zeta)+\bar{\mathcal{F}}(u,\bar\zeta)\,,
\end{equation}
where $\mathcal{F}(u,\zeta)$ is any function, holomorphic in~$\zeta$.

\subsubsection*{Kundt waves on type~D backgrounds}

Similarly, we can also describe gravitational waves propagating on general type D backgrounds discussed above. Indeed, the field equations for ${\lambda}$, ${\kappa}$, and ${g}$ are linear, and the equation for ${h}$ is linear in ${h}$ (with nonlinear terms with ${\lambda}$ and ${\kappa}$ as a `source'). We can thus easily superpose a pure gravitational-wave contribution of the form \eqref{vacuumwaves} on top of \emph{any} background metric function ${h}$, keeping the values of ${\lambda, \kappa}$, and ${g}$ unchanged. In particular, considering the exceptional Pleba\'{n}ski--Hacyan type~D background \eqref{genPlHac}, the family of gravitational waves described by \cite{GarciaAlvarez:1984} is obtained.

\subsubsection*{More general Kundt waves}\label{sc:genKundt}

The equation \eqref{vaclambdaeq} is the special Helmholtz equation on the transverse space such that the coefficient of the `mass' term is exactly given by the curvature of the transverse space. Its general solution can thus be parametrized by a single function ${\mathcal{L}(u,\zeta)}$, holomorphic in ${\zeta}$, as
\begin{equation}\label{lambdasol}
  \lambda = \mathcal{L}_{,\zeta} + \bar{\mathcal{L}}_{,\bar\zeta} 
    - 2\mathcal{L}\,\bigl(\log P\bigr)_{\!,\zeta} - 2\bar{\mathcal{L}}\,\bigl(\log P\bigr)_{\!,\bar\zeta}\;.
\end{equation}
Again, the functions ${\kappa}$ and ${g}$ can be gauged away, ${\kappa=g=0}$, and the electro-vacuum condition ${\iota=p=q=0}$ implies  ${\laplace\hat h=0}$, see~\eqref{heqpot}. However, now we have an additional contribution to ${h}$ thanks to a nontrivial ${\lambda}$, cf.~\eqref{hath}:
\begin{equation}\label{hsolgenwave}
  h= \mathcal{F}(u,\zeta)+\bar{\mathcal{F}}(u,\bar\zeta) + \frac12\Lambda_+ \lambda^2\;.
\end{equation}

We have thus obtained an explicit form of a general Kundt electro-vacuum spacetime \eqref{m1}. Apart from nontrivial $H$, these most general gravitational waves within our class also have nontrivial metric functions $a_i$, given as ${a_\zeta= -i\Lambda_+P^{-2}(\mathcal{L}+\bar{\mathcal{L}})}$. We are not aware of a discussion of such waves in the literature. 

It should, however, be mentioned that some of these solutions have unphysical behavior of the metric functions---typical solutions of \eqref{lambdasol} and \eqref{hsolgenwave} have singularities or diverge in transverse directions. They thus cannot be interpreted as globally well-behaved gravitational waves. Nevertheless, some of them can be interpreted as external vacuum solutions around a localized matter source, e.g., around a beam of null radiation or gyratonic matter. Such solutions will be discussed in the next section. Here we only note that they can be constructed from given matter sources using the Green functions. They are regular and satisfy vacuum equations outside the sources.

Since the `mass' term in \eqref{vaclambdaeq} has a special value, this equation also admits \emph{globally regular} solutions. Regular solutions for ${\lambda}$ are exactly those discussed for the type~D backgrounds, namely, given by \eqref{gentypeDkl}, \eqref{gentypeDPQ} (or \eqref{reglambdasph}). The solution for ${h}$ which leads to the regular geometry is given by \eqref{hsolgenwave} with sufficiently smooth ${\mathcal{F}}$, e.g., when it is quadratic in ${\zeta}$.

\subsection{Gyratons on the flat background}\label{ssc:flatgyr}

Our class of solutions also contains, as a subcase, the original gyraton on a flat background \cite{Bonnor:1970b:INTHP:}. Indeed, for a vanishing cosmological constant and  electromagnetic field absent, the background is Minkowski space. If we admit only an ${r}$ independent gyraton source (i.e., if we assume ${\div j=0}$, ${j_u=\iota}$) and if we employ the gauge ${g=0}$, we immediately obtain the solution discussed in \cite{Fro-Is-Zel:2005:PHYSR4:, Fro-Fur:2005:PHYSR4:}.

\section{Properties of the gyraton solutions}\label{sc:interpret}

\subsection{Character of the gyratons}\label{ssc:gyraton}

Now we concentrate on nontrivial gyratons contained in the above class. A characteristic feature of the gyratonic matter is a nonvanishing source ${j_i}$ in \eqref{m7} or, equivalently, its two potentials ${p}$ and ${q}$, cf.~\eqref{sourcepot}. The gyratonic matter moves with the speed of light, as can be identified by inspecting the dependence of the metric function on the coordinate ${u}$. From the form of the metric \eqref{m1} we infer that ${u}$ is a null coordinate, with null generators given by the principal null congruence ${\partial_r}$. All of the metric functions can depend on this coordinate, and this dependence is not restricted by the field equations. It means that the profile of the gyraton in the ${u}$ direction can be prescribed arbitrarily. Thanks to a trivial ${r}$ dependence of the fields, such a profile remains essentially unaltered (except for the ``cooling effect'' discussed below). This can be understood as a motion of the gyraton in the direction of the null congruence ${\partial_r}$.

The characteristic spatial components of the gyraton stress-energy tensor represent a possibility of an internal energy flow of otherwise null radiation. It can be naturally split into two components.

The divergence-free component, controlled by the source potential ${q}$, corresponds to a ``rotational'' part of the energy flow. However, since the gyratonic matter is null, the nature of the ``rotation'' must be internal---it describes a spin of the null fluid. This kind of the source was discussed in the context of the gyratons in flat spacetime \cite{Fro-Is-Zel:2005:PHYSR4:, Fro-Fur:2005:PHYSR4:} and in anti-de~Sitter space \cite{Fro-Zel:2005:PHYSR4:}. From the field equation \eqref{lambdaeq} we observe that this ``rotational'' part of the source gives rise to the component of the metric function ${a_i}$ determined by the potential ${\lambda}$ via \eqref{klpotdef}. This component is independent of the gauge, so the presence of the ``rotational'' part of the gyraton source necessary leads to the nondiagonal component ${a_i}$ in the metric ${\eqref{m1}}$.

The rotation-free component of the gyraton source, controlled by the potential ${p}$, has a different character. As can be read out from the conservation law \eqref{jdecomp}, the source with a nontrivial divergence ${\div j}$ describes an internal flow of the energy in the gyraton beam which changes its internal energy ${j_u}$ with ${r}$. We could thus understand the ${p}$ component of the source as some kind of ``cooling'' which steadily decreases the energy density of the gyraton beam. Such a kind of the energy transfer is not very plausible physically, mainly because the cooling should occur in matter moving with the speed of light. It inevitably leads to an unnatural causal behavior of the source.

Indeed, it is easy to check that the gyraton stress-energy tensor \eqref{m7} (composed by either a ${p}$ or a \mbox{${q}$ component}) does not satisfy neither a null, weak, strong nor dominant energy condition. However, for a spinning matter it is not so surprising---bad causal behavior is typical for spinning relativistic objects when they are idealized excessively. 

From the equation \eqref{kappagp} we also observe that the ${p}$ part of the source controls the combination ${g-\Lambda_-\kappa}$ of the metric functions. Splitting its influence between ${g}$ and ${\kappa}$ is just a matter of a gauge choice. We have already discussed that it is possible to eliminate either of them but not simultaneously. A gyraton source composed just from the rotation-free component (${p\neq0}$, ${q=0}$) thus does not necessary lead to a nondiagonal component ${a_i}$ in the metric---its influence can be gauged away entirely into the metric function ${g}$, and vice versa.

\subsection{Geometrical properties of the principal null
congruence}\label{ssc:prop}

Let us now briefly discuss geometrical properties of the gyraton solutions.
The additional property \eqref{aiassumption}, ${\partial_r a_i=0}$,
characteristic for the subclass of spacetimes discussed here, has a
consequence that the null vector ${k}$ is \emph{recurrent},
\cite{Prav-Prav:2002:CLAQG:, Step:2003:Cam:}, namely,
\begin{equation}\label{recurrentk}
k_{\alpha;\beta}=(-\partial_{r}H) k_{\alpha}k_{\beta}.
\end{equation}
The null character of $k$ and the condition \eqref{recurrentk}
also imply that the null congruence with tangent vector $k$ is
geodesic, expansion-free, sheer-free, and twist-free and thus belongs 
to the Kundt class. 

The condition \eqref{aiassumption} and the condition \eqref{Heq} 
(the function ${H}$ is at most quadratic in $r$, with
a constant coefficient in front of $r^2$) guarantee that these Kundt
metrics are of the CSI type \cite{Coley-Her-Pel:2006:CLAQG}. For these
metrics it was also shown that there exists a ($u$-dependent) 
diffeomorfhism $\tilde{x}^i=\tilde{x}^{i}(u,x^{k})$ such
that the transverse metric \eqref{trmetric} can be made $u$-independent. We have already used
this property at the very beginning when we applied the conditions
\eqref{udepofP}. Moreover, the transverse space is locally homogeneous. 

For the complete \mbox{four-dimensional} gyraton spacetime, 
it was demonstrated in \cite{Coley-Gib-Her-Pope:2008:CLAQG} that there 
always exists a related locally homogeneous spacetime 
which has invariants that are identical to those of the Kundt CSI metric. This
``background'' metric can be obtained by setting ${a_{i}=g=h=0}$, 
\begin{equation}\label{hom}
d s_{\mathrm{bg}}^2= d s_{\perp}^2 -2\,\grad u\,\grad r + \Lambda_- r^2 \grad u^2\;,
\end{equation}
which is exactly the metric for direct-product background spacetimes
\eqref{n3}.

The condition \eqref{aiassumption} is also equivalent to the fact that the 2-spaces orthogonal to the transverse spaces are surface-forming.

It is a general property of the Kundt family that the vector ${k}$ is the principal null direction of the spacetime. By determining its degeneracy we can thus identify the algebraic type. To proceed, it will be convenient to introduce an aligned complex null tetrad ${\{k,\,l,\,m,\,\mb\}}$. There exists a standard choice of such a tetrad in the context of the Kundt family of spacetimes \cite{Step:2003:Cam:}, namely
\begin{equation}\label{b-vectors}
\begin{aligned}
  k&=\partial_{r}\;,\\
  l&=\partial_{u}
     -P^2\bigl(a_{x}\partial_{x}+a_{y}\partial_{y}\bigr)
     -\bigl(H+{\textstyle\frac{1}{2}}a^2\bigr)\,\partial_{r}\;,\\
  m&=\frac{P}{\sqrt{2}}(\partial_{x}+i\partial_{y})\;,\\
  \mb&=\frac{P}{\sqrt{2}}(\partial_{x}-i\partial_{y})\;.
\end{aligned}
\end{equation}
The spacelike complex vectors ${m^a}$ and ${\mb^a}$ are tangent to the transverse space.
Clearly, ${[k,l]=-(\partial_rH)\,k}$, so the space spanned on ${k}$ and ${l}$ is indeed surface-forming.
The dual frame in the space of 1-forms reads
\begin{equation}\label{b-forms}
\begin{aligned}
\frm{k}&=\grad r + (H+{\textstyle\frac{1}{2}}a^2)\, \grad u\;,\\
\frm{l}&=\grad u\;,\\
\frm{m}&=\frac{1}{\sqrt{2}P}(\grad x{-}i\grad y)+\frac{P}{\sqrt{2}}(a_{x}{-}ia_{y})\,\grad u\;,\\
\frm{\mb}&=\frac{1}{\sqrt{2}P}(\grad x{+}i\grad y)+\frac{P}{\sqrt{2}}(a_{x}{+}ia_{y})\,\grad u\;.
\end{aligned}\end{equation}

Calculating the Newman--Penrose spin coefficients with respect to this tetrad (see the following section for the nontrivial ones), we recover again the general properties that the congruence is nonexpanding and nontwisting (${\NP\rho=0}$), sheer-free (${\NP\sigma=0}$), geodesic and affinely parametrized (${\NP\kappa=\NP\eps=0}$). In addition, from ${\NP\kappa=\NP\pi=\NP\eps=0}$, it follows that the tetrad \eqref{b-vectors} is parallelly transported along the null congruence. Moreover, the condition \eqref{aiassumption} is directly related to the vanishing coefficient ${\NP\tau=0}$.

\subsection{NP formalism in complex coordinates}\label{ssc:NP}

It turns out to be more convenient (and common in the literature on Kundt spacetimes) to introduce complex coordinates in the transverse space. Instead of conformally flat real coordinates ${x}$ and ${y}$ we will now use the complex coordinates ${\zeta}$ and ${\bar{\zeta}}$ such that
\begin{equation}\label{f1}
 \zeta=\frac{1}{\sqrt{2}}(x+iy)\;.
\end{equation}
The coordinate 1-forms and vector fields transform as
\begin{equation}\label{f2}
\begin{aligned}
  \grad \zeta&=\frac{1}{\sqrt{2}}(\grad x+i\grad y)\;,&
  \partial_{\zeta} &=\frac{1}{\sqrt{2}}(\partial_{x}-i\partial_{y})\;,\\
  \grad \bar{\zeta}&=\frac{1}{\sqrt{2}}(\grad x-i\grad y)\;,&
  \partial_{\bar\zeta}&=\frac{1}{\sqrt{2}}(\partial_{x}+i\partial_{y})\;,
\end{aligned}
\end{equation}
and the transverse Laplace operator \eqref{lapldef} on any scalar ${\psi}$ becomes
\begin{equation}\label{Delta}
  \laplace\psi = 2P^2\psi_{,\zeta\bar\zeta}\;.
\end{equation}

Instead of the real 1-form components ${a_i}$ it is customary to introduce a complex function ${W(u,\zeta,\bar\zeta)}$ by
\begin{equation}\label{f4}
W=-a_\zeta=-{\textstyle\frac{1}{\sqrt{2}}}(a_{x}{-}ia_{y})\;,\quad
\overline{W}=-a_{\bar\zeta}
\;.
\end{equation}
Substituting for ${a_i}$ the potentials via \eqref{klpotdef} and using
${\epsilon^\zeta{}_{\zeta}=-\epsilon^{\bar\zeta}{}_{\bar\zeta}=-i}$,
${\epsilon^\zeta{}_{\bar\zeta}=\epsilon^{\bar\zeta}{}_{\zeta}=0}$,
we find that
\begin{equation}\label{Wkl}
W=-(\kappa+i\lambda)_{,\zeta}\;,\quad
\overline{W}=-(\kappa-i\lambda)_{,\bar\zeta}\;.
\end{equation}
We also obtain
\begin{equation}\label{Wrels}
\begin{gathered}
a^2=2P^2W\overline{W}\;,\\
\laplace\kappa=\div a = - P^2 \bigl(\overline{W}_{\!,\zeta}+W_{\!,\bar{\zeta}}\bigr)\;,\\
\laplace\lambda=-\rot a = -b 
   = i P^2 f_{\zeta\bar\zeta} 
   = i P^2 \bigl({W}_{\!,\bar\zeta}-\overline{W}_{\!,\zeta}\bigr)\;.
\end{gathered}
\end{equation}

The metric \eqref{m1} in complex coordinates then reads
\begin{equation}\label{metriccx}
d s^2=
   \frac{2}{P^2}\grad \zeta \grad \bar{\zeta}-2\grad u\grad r-2H\grad u^2   -2\bigl(W\grad \zeta {+} \overline{W}\grad \bar{\zeta}\bigr) \grad u\;,
\end{equation}
the canonical form \eqref{P1} of ${P}$ is
\begin{equation}\label{f9}
P=1+\frac12{\Lambda_{+}}\zeta \bar{\zeta}\;,
\end{equation}
and the Maxwell tensor \eqref{EMF} takes form
\begin{equation}\begin{split}\label{EMFcx}
F&=E\,\grad r\wedge\grad u + B\, \frac{i}{P^2}\grad \zeta\wedge\grad \bar\zeta\\
 &\qquad\qquad+\sigma_{\zeta}\,\grad u\wedge\grad \zeta+\sigma_{\bar\zeta}\,\grad u\wedge\grad \bar\zeta\;,
\end{split}\end{equation}
where ${\sigma_\zeta=(\sigma_x-i\sigma_y)/\sqrt{2}}$.

The tetrads \eqref{b-vectors} and \eqref{b-forms} are closely related to the introduced complex coordinates:
\begin{equation}\label{vectors1}
\begin{aligned}
  k&=\partial_{r}\;,\\
  l&=\partial_{u}+P^2\bigl(W\partial_{\bar{\zeta}}+\overline{W}\partial_{\zeta}\bigr)
     -(H+P^2W\overline{W})\partial_{r}\;,\\
  m&=P\partial_{\bar{\zeta}},\\
  \mb&=P\partial_{\zeta},
\end{aligned}
\end{equation}
and
\begin{equation}\label{forms1}
\begin{aligned}
\frm{k}&=\grad r+(H+P^2W\overline{W}) \grad u\;,\\
\frm{l}&=\grad u\;,\\
\frm{m}&=\frac{1}{P}\,\grad \bar{\zeta}-PW\grad u\;,\\
\frm{\mb}&=\frac{1}{P}\,\grad \zeta-P\overline{W}\grad u\;.
\end{aligned}
\end{equation}
The list of nontrivial NP coefficients is then
\begin{equation}\label{f10}
\begin{aligned}
\NP\lambda&=(P^2W)_{,\zeta}\;,\\
\NP\mu&=\frac{1}{2}P^2\bigl(W_{,\bar{\zeta}}+\overline{W}_{,\zeta}\bigr)\;,\\
\NP\nu&=P(H+P^2W\overline{W})_{,\zeta}\;,\\
\NP\gamma&={\textstyle\frac{1}{2}}\Bigl[\partial_{r}H
   +{\textstyle\frac{1}{2}}\bigl((P^2\overline{W})_{,\zeta}-(P^2W)_{,\bar{\zeta}}\bigr)\Bigr]\;,\\
\NP\alpha&={\textstyle\frac{1}{2}}P_{,\zeta}\;,\\
\NP\beta&=-{\textstyle\frac{1}{2}}P_{,\bar{\zeta}}\;.
\end{aligned}
\end{equation}

The source equations can be recovered in the Newman--Penrose formalism by comparing the components of the Ricci tensor with the corresponding components of the electromagnetic and gyraton stress-energy tensor. The general form of nonvanishing Ricci scalars for the metric \eqref{metriccx} is listed in Appendix \ref{apx:NP} in equations \eqref{RicciPhi}. After substituting the potentials ${\kappa}$, ${\lambda}$, cf.\ equations~\eqref{Wkl}, and using the relations \eqref{f4}, \eqref{Wrels}, we find
\begin{equation}\label{RicciPhipot}
\begin{aligned}
\Phi_{11}&=\frac{1}{2}\rho\;,\\
\Phi_{12}&=\frac{P}{2}\Bigl[
   -2\rho\kappa
   +\bigl(g-\Lambda_{-}\kappa\bigr)
   +i\bigl(\frac{1}{2}\laplace\lambda+\Lambda_+\lambda\bigr)
   \Bigr]_{,\bar\zeta}\;,\\
\Phi_{22}&=\frac{1}{2}\Bigl[
   r\,\laplace\! (g-\Lambda_- \kappa)
   +iP^2\bigl(a_{\zeta} b_{,\bar{\zeta}}{-}a_{\bar{\zeta}} b_{,\zeta}\bigr)\\
   &\quad
   +\laplace h
   +\frac{1}{2}\,b^2
   +\Lambda_{+}a^2
   +g\laplace\!\kappa +\partial_u\laplace\!\kappa
   \Bigr]\;,\\
R&=24\NP\Lambda =4\Lambda\;.
\end{aligned}
\end{equation}
The constants ${\Lambda}$ and ${\rho}$ have entered these expressions via combinations ${\Lambda=\frac12(\Lambda_++\Lambda_-)}$ and ${\rho=\frac12(\Lambda_+-\Lambda_-)}$ of the constants ${\Lambda_\pm}$ which parametrize the metric \eqref{metriccx} through \eqref{Heq} and \eqref{Peq}. Their relation to the cosmological constant ${\Lambda}$ and the electromagnetic energy density ${\rho}$ is established by comparing these components to the cosmological term and to the corresponding components of the stress-energy tensors. For the electromagnetic field, the nonvanishing components are
\begin{equation}\label{EMTPhi}
\begin{aligned}
\Phi^\EM_{11}&=\frac{1}{2}\rho\;,\\
\Phi^\EM_{12}&=-P\rho\,\kappa_{,\bar\zeta}\;,\\
\Phi^\EM_{22}&=2P^2\rho\, \kappa_{,\zeta}\,\kappa_{,\bar\zeta}\;,
\end{aligned}
\end{equation}
with ${\rho}$ given by \eqref{rhodef}. Similarly, for the gyratonic matter we obtain
\begin{equation}\label{gyrTPhi}
\begin{aligned}
\Phi^\gyr_{11}&=0\;,\\
\Phi^\gyr_{12}&=\frac{P}{2}j_{\bar\zeta}\;,\\
\Phi^\gyr_{22}&=\frac{1}{2}j_u-P^2\bigl(a_{\bar\zeta}j_\zeta + a_{\zeta}j_{\bar\zeta}\bigr)\;.
\end{aligned}
\end{equation}
The first and second source equations \eqref{kappageq} and \eqref{heqpot} are obtained from the above components ${\Phi_{12}}$ and ${\Phi_{22}}$, respectively, by realizing that
\begin{equation}\label{sourcecx}
j_\zeta = (p+iq)_{,\zeta}\;,\quad
j_{\bar\zeta} = (p-iq)_{,\bar\zeta}\;.
\end{equation}

Finally, in Appendix \ref{apx:NP} we also present the Weyl scalars \eqref{WeylPsi} in the form which follows directly from the metric \eqref{metriccx} without using the field equations. If we introduce the potentials and employ the field equations, the nontrivial scalars reduce to
\begin{align}
 \Psi_{2}&=-\frac{1}{3}\Lambda\;,\notag\\
 \Psi_{3}&=P\Lambda\,\kappa_{,\zeta}+\frac{P}{2}(p-iq)_{,\zeta}\;,\label{WeylPsipot}\\
 \Psi_{4}&=
         r\bigl(P^2 (p-i\Lambda_- \lambda)_{,\zeta}\bigr)_{,\zeta}
         +\bigr(P^2\hat h_{,\zeta}\bigr)_{,\zeta}\notag\\
     &\quad
         +\bigl[\Lambda_+\lambda+i\Lambda_-\kappa+i p+i\partial_u\bigr]
         \bigl(P^2\lambda_{,\zeta}\bigr)_{,\zeta}\notag\\
     &\quad
         +2i P^2 q_{,\zeta} (\kappa+i\lambda)_{,\zeta}
         +p \bigl(P^2\kappa_{,\zeta}\bigr)_{,\zeta}
         -2\Lambda P^2 (\kappa_{,\zeta})^2\;,\notag
\end{align}
where ${\hat h}$ is given by \eqref{hath}.

Since ${\Psi_0=\Psi_1=0}$ and, for a nonvanishing cosmological constant, ${\Psi_{2}\neq0}$, we conclude that the vector ${k}$ is the double degenerate principal null direction and the gyraton spacetime is of the algebraical type II. The conditions for further algebraic degeneracy to the type D are, in the vacuum case, discussed in Section \ref{ssc:backgroundsTypeD} and in Appendix~ \ref{apx:backgroundsTypeD}. In the nonvacuum case these conditions are rather strong: for example, there are no nontrivial type~D gyratons with ${p=0}$.

For the vanishing cosmological constant ${\Lambda=0}$, the presence of a nontrivial rotational gyratonic matter (given by the potential ${q}$; terms with ${\kappa}$ and ${p}$ are not significant, as they can be cancelled by a suitable gauge) guarantees that the spacetime is of type III. The spacetime reduces to type N only for ${\Lambda=0}$ and ${q=0}$.

Comparing \eqref{RicciPhipot} with \eqref{WeylPsipot}, or directly \eqref{RicciPhi} with \eqref{WeylPsi}, we find that ${\Phi_{12}}$ and ${\Psi_3}$ are closely related, namely
\begin{equation}\label{f16}
\Phi_{12}+{\overline{\Psi}}_{3}=Pg_{,\bar{\zeta}}\,.
\end{equation}
The radiative characteristic of the gravitational field ${\Psi_3}$ is thus determined by the matter component ${\Phi_{12}}$, up to the term which can be controlled by the gauge. In the gauge ${g=0}$, we have directly ${\Psi_3=-\overline{\Phi}_{12}}$.

\subsection{NP components in the gauge invariant tetrad}\label{ssc:NPginv}

The tetrad \eqref{b-vectors} introduced in Section \ref{ssc:prop} is parallelly transported and transverse-surface forming, however it is not gauge invariant under the transformation \eqref{gauge}. Consequently, the spin coefficients and curvature scalars can have rather nontrivial dependence on the gauge transformation. It is possible to choose another tetrad which is gauge independent. Actually, it is a tetrad which is also well known in the context of the Kundt family of solutions \cite{Step:2003:Cam:}. This is related to the tetrad \eqref{b-vectors} by a null rotation with the vector ${k}$ fixed, i.e., it is also aligned with the principal null congruence. Explicitly, this null rotation is 
\begin{equation}\begin{aligned}\label{t1}
k'&=k\;,\\
m'&=m + K k\;,\\
l'&=l + K \mb + \overline{K} m + K\overline{K}  k\;,
\end{aligned}\end{equation}
with the complex parameter ${K=-P\overline{W}}$.
Thus, the gauge invariant tetrad of null vectors is
\begin{equation}\label{vectors11}
\begin{gathered}
k'=\partial_{r}\;,\quad
l'=\partial_{u}-H\partial_{r}\;,\\
m'=P(\partial_{\bar{\zeta}}-\overline{W}\partial_{r})\;,\quad
\mb'=P(\partial_{\zeta}-W\partial_{r})\;,
\end{gathered}
\end{equation}
and for the dual frame of 1-forms we obtain

\begin{gather}
\frm{k}{}'=\grad r+W\grad\zeta +\overline{W}\grad\bar{\zeta}+H\grad u\;,\quad
\frm{l}{}'=\grad u\;,\notag\\
\frm{m}{}'=\frac{1}{P}\grad\bar\zeta\;,\quad
\frm{\mb}{}'=\frac{1}{P}\grad\zeta\;.\label{forms11}
\end{gather}

The corresponding nontrivial spin coefficients are
\begin{equation}\begin{aligned}\label{sc10}
&\NP\lambda'=0\;,\\
&\NP\mu'=-\frac{i}{2}b\;,\\
&\NP\nu'=P\bigl(H_{,\zeta}-W\partial_{r}H -\partial_{u}W\bigr)\;,\\
&\NP\gamma'=\frac{1}{2}\bigl(\partial_{r}H-\frac12ib\bigr)\;,\\
&\NP\alpha'=\frac{1}{2}P_{,\zeta}\;,\\
&\NP\beta'=-\frac{1}{2}P_{,\bar{\zeta}}\;,
\end{aligned}\end{equation}
and the nonvanishing  Ricci scalars are
\begin{equation}\label{sc3}
\begin{aligned}
\Phi_{11}&=\frac{1}{2}\rho\;,\\
\Phi_{12}
&=\frac{P}{2}\Bigl[
   -2i\rho\lambda
   +\bigl(g-\Lambda_{-}\kappa\bigr)
   +i\bigl(\frac{1}{2}\laplace\lambda+\Lambda_+\lambda\bigr)
   \Bigr]_{,\bar\zeta}\;,\\
\Phi_{22}
&=\frac{1}{2}\Bigl[
   r\,\laplace\!(g-\Lambda_-\kappa)
   +2P^2\bigl(a_{\bar{\zeta}}g_{,\zeta}{+}a_{\zeta} g_{,\bar{\zeta}}\bigr)\\
   &\quad+\frac{1}{2}b^2
   +\laplace h
   -\Lambda_{-}a^2
   +g\,\laplace\kappa +\partial_u\laplace\kappa
   \Bigr]\;.\\
\end{aligned}
\end{equation}
For the electromagnetic field, the nonvanishing components are
\begin{equation}\label{EMTPhi2}
\begin{aligned}
\Phi^\EM_{11}&=\frac{1}{2}\rho\;,\\
\Phi^\EM_{12}&=-iP\rho\,\lambda_{,\bar\zeta}\;,\\
\Phi^\EM_{22}&=2P^2\rho\, \lambda_{,\zeta}\,\lambda_{,\bar\zeta}\;.
\end{aligned}
\end{equation}
Similarly, for the gyratonic matter we obtain
\begin{equation}\label{gyrTPhi2}
\begin{aligned}
\Phi^\gyr_{11}&=0\;,\\
\Phi^\gyr_{12}&=\frac{P}{2}j_{\bar\zeta}\;,\\
\Phi^\gyr_{22}&=\frac{1}{2}j_u\;.
\end{aligned}
\end{equation}
The nonvanishing Weyl scalars read
\begin{align}
 \Psi_{2}&=-\frac{1}{3}\Lambda\;,\notag\\
 \Psi_{3}&=-iP\Lambda\lambda_{,\zeta}+\frac{P}{2}(p-iq)_{,\zeta}\;,\notag\\
 \Psi_{4}&=
         r\bigl(P^2 (p-i\Lambda_- \lambda)_{,\zeta}\bigr)_{,\zeta}
         +\bigr(P^2\hat h_{,\zeta}\bigr)_{,\zeta}\label{sc3a}\\
     &\quad
         +\bigl[\Lambda_-\lambda+i\Lambda_-\kappa+i p+i\partial_u\bigr]
         \bigl(P^2\lambda_{,\zeta}\bigr)_{,\zeta}\notag\\
     &\quad
         +2P^2 p_{,\zeta} (\kappa+i\lambda)_{,\zeta}
         +p \bigl(P^2\kappa_{,\zeta}\bigr)_{,\zeta}
         +2\Lambda_{-}P^2 (\lambda_{,\zeta})^2\;.\notag
\end{align}

\subsection{Electromagnetic field}\label{ssc:elmag}

In our ansatz made in Section \ref{ssc:ansatz} we allowed the spacetime to be filled with the electromagnetic field \eqref{EMF}. This field does not have its own dynamical degrees of freedom---it is specified just by two constants ${E}$ and ${B}$. In the presence of a gyraton, this electromagnetic field is modified through the ${\grad u\wedge \sigma_i\grad x^i}$ terms. However, the transverse 1-form ${\sigma_i}$ is uniquely determined by the gyraton, see equations \eqref{phipot}, \eqref{phisol}.

The Maxwell tensor \eqref{EMF} can be split into two parts
\begin{equation}\label{EMFsplit}
 F = E \bigl(\grad r\wedge\grad u+\grad u \wedge \grad\kappa\bigr)
    +B \bigl(P^{-2}\grad x\wedge\grad y+\grad u \wedge \grad\lambda\bigr)\;.
\end{equation}
It is interesting to observe that the 2-form proportional to the constant ${B}$ is the four-dimensional Hodge dual of the 2-form proportional to ${E}$. Thus, the Maxwell tensor has a familiar structure of a linear combination of dual `electric' and `magnetic' parts. Moreover, after substituting \eqref{phisol} for ${\varphi}$, the field equations depend only on the `weights' ${E}$ and ${B}$ of the electric and magnetic parts through the constant ${\rho=\frac{\varkappa\epso}{2}(E^2+B^2)}$ (via the constants ${\Lambda_\pm=\Lambda\pm\rho}$). The geometry of the spacetimes thus does not depend on a particular splitting of the electromagnetic field.

To inspect the algebraic structure of the electromagnetic field, we need the tetrad components ${\Phi_A}$ of the Maxwell tensor. With respect to the parallelly transported tetrad \eqref{vectors1}, we obtain
\begin{equation}\begin{aligned}\label{EMPhi}
&\Phi_{0}=0\;,\\
&\Phi_{1}=\frac{1}{2}(E+iB)\;,\\
&\Phi_{2}=-P(E+iB)\,\kappa_{,\zeta}\;,
\end{aligned}\end{equation}
while with respect to the gauge invariant tetrad \eqref{vectors11} we get\begin{equation}\begin{aligned}\label{EMPhiginv}
&\Phi_{0}=0\;,\\
&\Phi_{1}=\frac{1}{2}(E+iB)\;,\\
&\Phi_{2}=iP(E+iB)\,\lambda_{,\zeta}\;.
\end{aligned}\end{equation}
It follows that the electromagnetic field is aligned with the principal null direction ${k}$ of the gravitation field, but this vector is not a double degenerate vector of the field.

The corresponding tetrad components of the electromagnetic stress-energy tensor ${\Phi^\EM_{AB}=\varkappa\epso\Phi_A{\bar\Phi}_B}$ have been listed in \eqref{EMTPhi} and \eqref{EMTPhi2}. Notice that the ${\Phi^\EM_{12}}$ and ${\Phi^\EM_{22}}$ components of \eqref{EMTPhi} can be simultaneously canceled by the gauge choice ${\kappa=0}$. This choice also cancels the component ${\Phi_2}$ in \eqref{EMPhi}.

\section{Green functions}\label{sc:GreenFc}

In Section \ref{sc:gyreq} we demonstrated that for our ansatz the Einstein--Maxwell equations effectively reduce to the Poisson equations 
\begin{equation}\label{poisson}
  \laplace\psi = -s\;,
\end{equation}
(e.g., equations \eqref{lambdab}, \eqref{heqpot} for ${\lambda}$ and ${\hat h}$), and to the Helmholtz--Poisson equations
\begin{equation}\label{helmholtz}
  \laplace\psi + R_\perp\psi = -s\;,
\end{equation} 
(equations \eqref{lambdaeq} and \eqref{rotjeq} for ${\lambda}$ or ${b}$). These equations on the two-dimensional transverse space can be solved using the Green functions ${G_{(0)}}$ and ${G_{(1)}}$, respectively. Such functions satisfy\footnote{%
In the case when there exist normalizable zero modes, one has to subtract a projector to the space of these modes from the delta function on the right-hand side. See a discussion in Appendix~\ref{apx:Green}.}  
\begin{gather}
  \laplace G_{(0)}(x,x') = -\delta(x,x')\;,\\
  [ \laplace + R_\perp ] G_{(1)}(x,x') = -\delta(x,x')\;, 
\end{gather} 
where ${x}$ and ${x'}$ are points in the transverse space.
The solutions are then given by the integral over the corresponding sources
\begin{equation}
\psi(x)=\int G_{(\nu)}(x,x')\,s(x')\,\sqrt{g'_{\perp}}\,d ^2{x'}\;.
\end{equation}
In particular,
\begin{equation}\begin{aligned}\label{grfc0}
\lambda(u,x)&=\int G_{(0)}(x,x')\,b(u,x')\,\sqrt{g'_{\perp}}\,d ^2{x'}\;,\\
\hat h(u,x)&=\int G_{(0)}(x,x')\,\chi(u,x')\,\sqrt{g'_{\perp}}\,d ^2{x'}\;,
\end{aligned}\end{equation}
with the source 
\begin{equation}\label{m41}
\chi = -\iota - q \,\laplace\lambda+ p \,\laplace\kappa +2 a^i p_{,i}\;,
\end{equation}
and
\begin{equation}\begin{aligned}\label{grfc1}
\lambda(u,x)&= 2\int G_{(1)}(x,x')\,q(u,x')\,\sqrt{g'_{\perp}}\,d ^2{x'}\;,\\
b(u,x)&=       2\int G_{(1)}(x,x')\,\rot j(u,x')\,\sqrt{g'_{\perp}}\,d ^2{x'}\;.
\end{aligned}\end{equation}

It follows from the Einstein equations that the two-dimensional transverse space is a maximally symmetric space of constant curvature ${R_\perp = 2\Lambda_+}$, i.e., a plane ${E^2}$ for ${\Lambda_+=0}$, a sphere ${S^2}$ for ${\Lambda_+>0}$, and a hyperboloid ${H^2}$ (Lobachevsky plane) for ${\Lambda_+<0}$. The corresponding Green functions are known explicitly (see, e.g., \cite{Zelnikov:2008:JHEP:}) and they are discussed in more detail in  Appendix \ref{apx:Green}. Here we present only those results which are important for solution of our problem.

\subsection{Green functions for 2-plane $E^2$}\label{sc:plane}

For ${\Lambda_+\!\!=0}$ the transverse space is the flat plane. Both the Green functions coincide and they have the form
\begin{equation}\label{FrFcE2}
  G_{(\nu)}(x,x') = -\frac{1}{2\pi}\,\,\log\ell(x,x')\;.
\end{equation}
Here ${\ell(x,x')}$ is the distance between the points ${x}$ and~${x'}$.
\subsection{Green functions for 2-hyperboloid $H^2$}\label{sc:hyper}

For ${\Lambda_{+}=-1/L^2 <0}$ the transverse space \eqref{trmetric} is hyperboloid of a constant negative curvature, $L$ being the curvature radius. It can be parametrized by different useful coordinate systems. Here, we list some of them which are frequently used in the literature, namely, hyperspherical, Poincar\'e, Lobachevsky, and projective coordinates, respectively:
\begin{equation}\begin{aligned}\label{h1}
\grad s_{\perp}^2
&=L^2\left(\,\grad \rho^2+\sinh^2\rho~\grad \phi^2\right)\\
&={\frac{L^2}{z^2}}(\,\grad t^2+\grad z^2)\\
&={L^2}\left(\,\grad \mu^2+\cosh^2\mu~\grad \tau^2\right)\\
&={\frac{1}{\left(1{+}\frac14{\Lambda_{+}}(x^2{+}y^2)\right)^2}}\,\bigl(\,\grad x^2+\grad y^2\bigr)\;.
\end{aligned}\end{equation}
Relations of the coordinates to the projective ones are
\begin{equation}\label{lobcoors}
\begin{aligned}
  x&=2L \tanh\frac\rho2\cos\phi\;,&
  y&=2L \tanh\frac\rho2\sin\phi\;,\\
  x&=L\frac{4t}{t^2+(1+z)^2}\;,&
  y&=L\frac{t^2+z^2-1}{t^2+(1+z)^2}\;,\\
  x&=\frac{2L\cosh\mu\sinh\tau}{\cosh\mu\cosh\tau+1}\;,&
  y&=\frac{2L\sinh\mu}{\cosh\mu\cosh\tau+1}\;.
\end{aligned}
\end{equation}
Clearly, the conformally flat coordinates used in the text correspond to the projective coordinates with the choice \eqref{P1} and to the Poincar\'e coordinates with the choice \eqref{P2}.

Because of the maximal symmetry of the space, the Green functions can be expressed only in terms of the geodesic distance between the points ${\ell(x,x')}$ or, more conveniently, of its function
\begin{equation}\begin{aligned}\label{h2}
\eta(x,x')=\cosh{\left(\sqrt{-\Lambda_{+}}\,\ell(x,x')\right)}\;.
\end{aligned}\end{equation}
The function ${\eta}$ in an explicit form reads
\begin{equation}\begin{aligned}\label{h3}
\eta
&=\cosh\rho\cosh\rho'-\sinh\rho\sinh\rho'\cos(\phi-\phi')\\
&=1+{(t-t')^2+\frac{(z-z')^2}{2zz'}}\\
&=\cosh\mu\cosh\mu'\cosh(\tau-\tau')-\sinh\mu\sinh\mu'\\
&=\frac{\left(1{-}\frac{\Lambda_{+}(x^2{+}y^2)}4\right)\left(1{-}\frac{\Lambda_{+}(x'^2{+}y'^2)}4\right)+\Lambda_{+}(xx'{+}yy')}
  {\left(1{+}\frac{\Lambda_{+}(x^2{+}y^2)}4\right)\left(1{+}\frac{\Lambda_{+}(x'^2{+}y'^2)}4\right)}\;.
\end{aligned}\end{equation}
Clearly, $\eta\in [1,\infty)$, with $\eta=1$ corresponding to coincident points, and ${\eta\to\infty}$ to an infinite distance. 
Using these quantities, the Green functions in question are
\begin{equation}\begin{aligned}\label{h4}
G_{(0)}({x,x'})&=-{\frac1{4\pi}}\log\left(\eta-\frac1\eta+1\right)\;,\\G_{(1)}({x,x'})&=-{\frac1{4\pi}}\left(\eta\log\left(\eta-\frac1\eta+1\right)+2\right)\;.
\end{aligned}\end{equation}
When $\eta\rightarrow\infty$, the Green functions tend to zero.

\subsection{Green functions for 2-sphere $S^2$}\label{sc:sphere}

When ${\Lambda_{+}=1/L^2>0}$ the metric \eqref{trmetric} describes a sphere
\begin{equation}\label{s1}
ds_{\perp}^2
={L^2}\bigl(\grad \theta^2\!+\sin^2\theta\grad \phi^2\bigr)
=\frac{\grad x^2\!+\grad y^2}{\left(1{+}\frac14{\Lambda_{+}(x^2\!+y^2)}\right)^2}\;.
\end{equation}
Spherical coordinates $(\theta,\phi)$  are related to the projective coordinates $(x,y)$, used in \eqref{m1} with ${P}$ given by \eqref{P1}, via the coordinate transformation
\begin{equation}\label{s2}
x=2L\tan\frac\theta2\,\cos\phi\;,\quad
y=2L\tan\frac\theta2\,\sin\phi\;.
\end{equation}
Then the Green functions are functions of
\begin{equation}\label{s3}
\eta({x,x'})=\cos{\bigl(\sqrt{\Lambda_+}\,\ell({x,x'})\bigr)}\;.
\end{equation}
Here $\eta$ varies in the interval $[-1,1]$ and has the form
\begin{equation}\begin{aligned}\label{s4}
\eta&=\cos\theta \cos\theta'+\sin\theta\sin\theta'\cos(\phi-\phi')\\
&=\frac{\left(1{-}\frac{\Lambda_{+}(x^2{+}y^2)}4\right)\left(1{-}\frac{\Lambda_{+}(x'^2{+}y'^2)}4\right)+\Lambda_{+}(xx'{+}yy')}
  {\left(1{+}\frac{\Lambda_{+}(x^2{+}y^2)}4\right)\left(1{+}\frac{\Lambda_{+}(x'^2{+}y'^2)}4\right)}\;.
\end{aligned}\end{equation}
The generic solution for the Green functions on a sphere is a linear combination of the Legendre functions ${\rm Q}_{\nu}(\eta)$ and ${\rm P}_{\nu}(\eta)$. However, the requirement of regularity at the antipodal point $\eta=-1$  singles out their particular combination. Also, one has to be cautious since both equations \eqref{poisson} and \eqref{helmholtz} on the compact sphere have normalizable zero modes---see Appendix \ref{apx:Green} for more details. Eventually, the Green functions read
\begin{equation}\begin{aligned}\label{s5}
G_{(0)}({ x,x'})&=-{\frac1{4\pi}}\log\left(1-\eta\right)\;,\\
G_{(1)}({ x,x'})&=-{\frac1{4\pi}}\Bigl(\eta\log\left({1-\eta}\right)+1\Bigr)\;.
\end{aligned}\end{equation}

Moreover, the left-hand side of \eqref{heqpot} is the Laplacian defined on a compact sphere. The integral of the Laplacian over the sphere has to be zero. This property imposes an integral condition on physically acceptable distributions of the stress-energy tensor, namely
\begin{equation}\begin{aligned}\label{s6}
\int_{S^2} \chi\,\sqrt{g_{\perp}}\,{d^2x}=0\;.
\end{aligned}\end{equation}
Similarly, for the equation \eqref{rotjeq} we also get an integral constraint
\begin{equation}\begin{aligned}\label{s7}
\int_{S^2} \cos\theta\,\,\rot j\,\sqrt{g_{\perp}}\,{d^2x}=0\;.
\end{aligned}\end{equation}
Because of this property the zero modes do not contribute to the components $a_i$ of the metric.
The constraints \eqref{s6} and \eqref{s7} appear only because $S^2$ is compact and are analogous to the property that closed worlds must have zero total energy, charge, or angular momentum \cite{Mar-Fro:1970:TMF:,BicakKrtous:2001}.

\section{Conclusion}\label{sc:conclusion}

We presented a new class of gyraton solutions on electro-vacuum background spacetimes which are formed by a direct product of two constant-curvature 2-spaces. These involve the (anti-)Nariai, Bertotti--Robinson, and Pleba\'{n}ski--Hacyan spacetimes in four dimensions. The background geometries are solutions of the Einstein--Maxwell equations corresponding to the uniform background electric and magnetic fields. The gyraton solutions are of Petrov type II and belong to the Kundt family of shear-free and twist-free nonexpanding spacetimes. 

Gyratons describe the gravitational field created by a stress-energy tensor of a spinning (circularly polarized) high-frequency beam of electromagnetic radiation, neutrino, or any other massless fields. They also provide a good approximation for the gravitational field of a beam of ultrarelativistic particles with a spin. The gyratons generalize standard {\it pp\,}-waves or Kundt waves by admitting a nonzero angular momentum of the source. This leads to other nontrivial components of the Einstein equations, namely ${G_{ui}+\Lambda g_{ui}=\varkappa T_{ui}}$, in addition to the pure radiation $uu$-component which appears for {\it pp\,}-waves or Kundt waves.

We have shown that all of the Einstein--Maxwell equations can be solved exactly  for any distribution of the matter sources (see Section \ref{sc:gyrsol} for a summary), and the problem has been reduced to finding the scalar Green functions on a two-dimensional sphere, plane or hyperboloid. These Green functions have been presented in detail in Section \ref{sc:GreenFc}.
Special cases of these gyraton solutions and their properties are discussed in Sections \ref{sc:knownsol} and \ref{sc:interpret}.

We have also studied the gyraton solutions using the Newman--Penrose formalism. The characteristic  term $a_{i}$, describing the rotational part of the gyraton, generates the nontrivial Ricci $\Phi_{12}$ and Weyl $\Psi_{3}$ scalars, in addition to the case of pure {\it pp\,} and Kundt waves. Curiously, there exists a very simple relation \eqref{f16} between them.

To complete our investigation, we have also studied gyratons on more general type-D backgrounds (including the exceptional Pleba\'{n}ski--Hacyan spacetime) which are not direct-product spaces. In addition, in Section \ref{sc:genKundt} we have identified a special subclass of the gyraton solutions---general vacuum Kundt waves which also contain cases previously not discussed in the literature.

A natural next step would be the study of gyratons in a full family of Kundt spacetimes, especially on conformally flat backgrounds, including the (anti-)de~Sitter universe. Another generalization could be their extension to higher dimensions, where, however, one has to deal with a richer possible structure of the transverse geometries.

\begin{acknowledgments}
H.~K. was supported by Grant No.~GA\v{C}R-205/09/H033 and by the Czech Ministry of Education under Project No.~LC06014. A.~Z. was financially supported by the Killam Trust and partly by the Natural Sciences and Engineering Research Council of Canada. P.~K. was supported by Grant No.~GA\v{C}R-202/09/0772, and J.~P. by Grant No.~GA\v{C}R-202/08/0187.
H.~K., P.~K., and J.P. thank the University of Alberta (Edmonton, Canada) for the hospitality
during their stays where this work has started and has been finished. A.~Z. is grateful to the Charles University (Prague, Czech Republic) for hospitality during his work on this paper. The authors are also grateful to Valeri Frolov and Dmitri Pogosyan for stimulating discussions.
\end{acknowledgments}

\appendix

\section{The Einstein equations}\label{apx:AppA}

Here we present geometric quantities which appear in the Einstein field equations, namely the Einstein tensor of the general metric \eqref{m1} and the electromagnetic stress-energy tensor corresponding to the field \eqref{EMF}.

The inverse to the metric \eqref{m1} is
\begin{equation}
\begin{split}\label{a0}
g^{\mu\nu}&\partial_{\mu}\partial_{\nu}=
  P^2\bigl(\partial_{x}\partial_{x}+\partial_{y}\partial_{y}\bigr)
  -2\partial_{u}\partial_{r}\\
  &+2P^2\bigl(a_{x}\partial_{x}+a_{y}P^2\partial_{y}\bigr)\partial_{r}
  +2(H+\frac{1}{2} a^2)\,\partial_{r}\partial_{r}.
\end{split}\raisetag{8ex}
\end{equation}

The stress-energy tensor of the electromagnetic field $T^\EM$,
according to the definition
\begin{equation}\label{defem}
T^\EM_{\mu\nu}=\epso\Bigl(
   F_{\mu\kappa}F_{\nu\lambda}\,g^{\kappa\lambda}
   -\frac{1}{4}g_{\mu\nu}\,F_{\kappa\lambda}F^{\kappa\lambda}
   \Bigr)\;,
\end{equation}
has nonzero components
\begin{align}
\varkappa T^\EM_{ur}&=\rho\;,\notag\\
\varkappa T^\EM_{uu}&=2H\rho+
   \varkappa\epso(\sigma-E a)^2\;,\notag\\
\varkappa T^\EM_{ux}&=\varkappa\epso\Bigl(\frac12(E^2{-}B^2)\,a_{x}-EB\, a_{y}
   -E\sigma_{x}+B \sigma_{y}\Bigr)\;,\notag\\
\varkappa T^\EM_{uy}&=\varkappa\epso\Bigl(\frac12(E^2{-}B^2)\,a_{y}+EB a_{x}
   -E\sigma_{y}-B \sigma_{x}\Bigr)\;,\notag\\
\varkappa T^\EM_{xx}&=\frac{\rho}{P^2}\;,\label{EMT}\\
\varkappa T^\EM_{yy}&=\frac{\rho}{P^2}\;,\notag
\end{align}
where the density ${\rho=\frac{\varkappa\epso}{2}(E^2{+}B^2)}$ was defined in~\eqref{rhodef}.

The Einstein tensor for the metric \eqref{m1} reads
\begin{equation}\label{EinsteinT}
\begin{aligned}
G_{ur}&= \laplace\!\log P\;,\\
G_{uu}&=\frac{1}{2}b^2+\laplace H +(\partial^2_{r}H)a^2+2a^i\partial_{r}H_{,i}\\
&\quad +(\partial_r H)\,\div a+\partial_{u}\div a
+2H\laplace\!\log P\;,\\
G_{ux}&=\frac{1}{2}b_{,y}-a_{x}\bigl(\laplace\!\log P-\partial^2_{r}H\bigr)+\partial_{r}H_{,x}\;,\\
G_{uy}&=-\frac{1}{2}b_{,x}-a_{y}\bigl(\laplace\!\log P-\partial^2_{r}H\bigr)+\partial_{r}H_{,y}\;,\\
G_{xx}&=\frac{1}{P^2}\,\partial^2_{r}H\;,\\
G_{yy}&=\frac{1}{P^2}\,\partial^2_{r}H\;,
\end{aligned}
\end{equation}
where
\begin{equation}\label{lapllogpcx}
  \laplace\!\log P=P\bigl(P_{,xx}+P_{,yy}\bigr)-\bigl(P_{,x}^2+P_{,y}^2\bigr)\;,
\end{equation}
cf.~\eqref{trsccurv}.
Here we have used only the metric \eqref{m1}, without assuming any other information about the metric functions. In particular, we have not used the field equations. To be more precise, in the components ${G_{ui}}$ and ${G_{ij}}$ we employed the fact that ${P}$ is ${u}$-independent; cf.\ relation \eqref{udepofP}. However, as we already mentioned in section \ref{ssc:transsp}, such a choice is always possible provided that the transverse spaces have the same homogeneous geometry---which can be derived just from the component ${G_{ur}}$.

\section{Newman--Penrose quantities}\label{apx:NP}

In this appendix we present the Ricci and Weyl curvature scalars corresponding to our subclass of the Kundt family given by the metric \eqref{metriccx}, 
\begin{equation}
d s^2=
   \frac{2}{P^2}\grad \zeta \grad \bar{\zeta}-2\grad u\grad r-2H\grad u^2   -2\bigl(W\grad \zeta {+} \overline{W}\grad \bar{\zeta}\bigr) \grad u\;,
\end{equation}
with functions ${W(u,\zeta,\bar\zeta)}$ and ${P(\zeta,\bar\zeta)}$ independent of ${r}$, or ${r}$ and ${u}$, respectively. The expressions below have not employed any field equations.

\subsection*{The standard tetrad}
With respect to the tetrad \eqref{vectors1} the only nonvanishing  Ricci scalars are
\begin{equation}\label{RicciPhi}
\begin{aligned}
\Phi_{11}&=\frac{1}{4}(\partial^2_{r}H+\laplace\!\log P)\;,\\
\Phi_{12}
&=\frac{P}{2}\biggl[\partial_{r}H_{,\bar{\zeta}}
  +(\laplace\!\log P)\overline{W}
  -\frac{1}{2}i b_{,\bar{\zeta}}\biggr]\;,\\
\Phi_{22}
&=\frac{P^2}{2}\biggl[\frac{b^2}{2P^2}
  +2H_{,\zeta\bar{\zeta}}
  -i(W b_{,\bar{\zeta}}{-}\overline{W} b_{,\zeta})\\
  &\quad+2(\laplace\!\log P) W\overline{W}
  -(\partial_{r}H\!+\partial_{u})(\overline{W}_{,\zeta}{+}{W}_{,\bar{\zeta}})\biggr]\;,\\
R&=24{\Lambda}_{\rm NP}=2(-\partial^2_{r}H+\laplace\!\log P)\;,
\end{aligned}
\end{equation}
and nonvanishing Weyl scalars read
\begin{equation}
\begin{aligned}\label{WeylPsi}
\Psi_{2}&=\frac{1}{6}(\partial^2_{r}H-\laplace\!\log P)\;,\\
\Psi_{3}
 &=\frac{P}{2}\left[\partial_{r}H_{,\zeta}-(\laplace\!\log P)W-\frac{1}{2}\,i\,b_{,\zeta}\right]\;,\\
\Psi_{4}
&=-iP^2W b_{,\zeta}-(\laplace\!\log P)P^2W^2-(P^2\partial_{u}W)_{,\zeta}\\
  &\quad+(P^2H_{,\zeta})_{,\zeta}-(\partial_{r}H)(P^2W)_{,\zeta}\;.\\
\end{aligned}
\end{equation}
Here, ${\laplace\!\log P}$ is given in \eqref{lapllogpcx}, and ${b=iP^2(\overline{W}_{,\zeta}-W_{,\bar\zeta})}$ as in \eqref{Wrels}.
The components of the electromagnetic field are 
\begin{equation}\begin{aligned}\label{f10a}
&\Phi_{0}=0,\\
&\Phi_{1}=\frac{1}{2}(E+iB),\\
&\Phi_{2}=P\left(-\sigma_{\zeta}+iBW\right)\;,
\end{aligned}\end{equation}
where ${\sigma_{\zeta}=\frac{1}{\sqrt{2}}(\sigma_{x}-i\sigma_{y})}$. 

\subsection*{The gauge invariant tetrad}

With respect to the gauge invariant tetrad \eqref{vectors11} we obtain that the only nonvanishing  Ricci scalars are
\begin{equation}\begin{aligned}\label{ff1}
\Phi_{11}&=\frac{1}{4}(\partial^2_{r}H+\laplace\!\log P)\;,\\
\Phi_{12}
  &=\frac{P}{2}\left[-\overline{W}(\partial^2_{r}H)
    +\partial_{r}H_{,\bar{\zeta}}
    -\frac{1}{2}i b_{,\bar{\zeta}}\right]\;,\\
\Phi_{22}
  &=\frac{P^2}{2}\biggl[\frac{b^2}{2P^2}+2H_{,\zeta\bar{\zeta}}
    -2(\overline{W}\partial_{r}H_{,\zeta}{+}W\partial_{r}H_{,\bar{\zeta}})\\
    &\quad+2(\partial^2_{r}H)W\overline{W}
    -(\partial_{r}H\!+\partial_{u})(\overline{W}_{\!,\zeta}{+}W_{\!,\bar{\zeta}})\biggr]\;,\\
\end{aligned}\end{equation}
the nonvanishing Weyl scalars read
\begin{equation}\begin{aligned}\label{ff2}
\Psi_{2}&=\frac{1}{6}(\partial^2_{r}H-\laplace\!\log P)\;,\\
\Psi_{3}
  &=\frac{P}{2}\left[-W(\partial^2_{r}H)+\partial_{r}H_{,\zeta}
    -\frac{1}{2}ib_{,\zeta}\right]\;,\\
\Psi_{4}
  &=P^2W^2(\partial^2_{r}H)-(P^2\partial_{u}W)_{,\zeta}+(P^2H_{,\zeta})_{,\zeta}\\
    &\quad-P^2 W (\partial_{r}H)_{,\zeta}
    -(P^2W\partial_{r}H)_{,\zeta}\;,\\
\end{aligned}\end{equation}
and, finally, the scalars for electromagnetic field are
\begin{equation}\begin{aligned}\label{sc2}
&\Phi_{0}=0\;,\\
&\Phi_{1}=\frac{1}{2}(E+iB)\;,\\
&\Phi_{2}=-P\left(EW+\sigma_{\zeta}\right)\;.
\end{aligned}\end{equation}

\section{All spacetimes of type D}\label{apx:backgroundsTypeD}

The purpose of this appendix is to derive all electro-vacuum solutions of the algebraic type~D within the class considered. Inspecting the field equations \mbox{\eqref{sumgkpeq}--\eqref{sumheq}} without the gyratonic matter, ${\iota=p=q=0}$, we find
\begin{gather}
g= \Lambda_-\kappa\;,\\
P^2\lambda_{,\zeta\bar\zeta}+\Lambda_+\lambda=0\;,\label{vaclambda}\\
P^2\hat h_{,\zeta\bar\zeta}=0\;,\label{vach}
\end{gather}
with ${\hat h}$ given by \eqref{sumhhatdef}. Using the gauge transformation \eqref{gauge}, we could eliminate both ${\kappa}$ and ${g}$, but this is not necessary in the following.

From \eqref{WeylPsipot} we infer that the vector ${k=\partial_r}$ is a double degenerate principal null direction. When ${\Psi_2\neq0}$, i.e., for a nonvanishing cosmological constant, the condition that there exists another degenerate null direction is
\begin{equation}
3\Psi_4\Psi_2=2\Psi_3^2\;,
\end{equation}
cf.~\cite{Step:2003:Cam:}. This reduces to the relation
\begin{equation}
\begin{split}
0&=i r \Lambda_- \bigl(P^2\lambda_{,\zeta}\bigr)_{,\zeta}\\
 &\quad +\bigl(P^2 \hat h_{,\zeta}\bigr)_{,\zeta}
  +\bigl[\Lambda_+\lambda+\Lambda_-\kappa+i\partial_u\bigr] \bigl(P^2\lambda_{,\zeta}\bigr)_{,\zeta}
  \;.
\end{split}
\end{equation}
Taking into account the ${r}$ dependence, we obtain the following two conditions\footnote{The case ${\Lambda_-=0}$ can be easily discussed separately.}:
\begin{gather}
\bigl(P^2\lambda_{,\zeta}\bigr)_{,\zeta}=0\;,\label{typeDlambda}\\
\bigl(P^2\hat h_{,\zeta}\bigr)_{,\zeta}=0\;.\label{typeDh}
\end{gather}
These equations must be accompanied by the condition \eqref{Peq} for the metric function ${P}$, namely,
\begin{equation}
P^2(\log P^2)_{,\zeta\bar\zeta}=\Lambda_+\;.\label{Peqapx}
\end{equation}

\subsection*{The case ${\Lambda_+\neq0}$}

Integrating \eqref{vaclambda}, we obtain
\begin{equation}\label{lambda,zeta}
\lambda_{,\zeta}=\frac{\bar{\mathcal{L}}}{P^2}\;,
\end{equation}
where the arbitrary function ${\bar{\mathcal{L}}(u,\bar\zeta)}$ can depend only on ${\bar\zeta}$ and ${u}$. First assuming ${\Lambda_+\neq0}$, we can substitute the ${\bar\zeta}$ derivative of \eqref{lambda,zeta} into \eqref{vaclambda}, which leads to
\begin{equation}\label{typeDlambdasol}
\lambda=\frac{\bar{\mathcal{L}}\,P^2{}_{,\bar\zeta}-\bar{\mathcal{L}}_{,\bar\zeta}\,P^2}{\Lambda_+ P^2}\;.
\end{equation}
Now we have to check the consistency of this general solution for ${\lambda}$ with \eqref{lambda,zeta} (since we have used the equation \eqref{vaclambda} to obtain ${\lambda}$). It turns out that the solution is consistent only thanks to the fact that the constants ${\Lambda_+}$ which appears in \eqref{vaclambda} and \eqref{Peqapx} are the same.

However, the function ${\bar{\mathcal{L}}}$ in \eqref{typeDlambdasol} is not arbitrary. The last condition which must be satisfied is that ${\lambda}$ is real. It is not straightforward to find the consequences for ${\bar{\mathcal{L}}}$ explicitly in a general case. But we can use the freedom in transverse diffeomorfhism to transform the solution of \eqref{Peqapx} into a particular form. It will be useful to use such transverse coordinates for which ${P}$ is linear  both in ${\zeta}$ and~${\bar\zeta}$: 
\begin{equation}\label{Plincond}
P_{,\zeta\zeta}=0\;.
\end{equation}
Explicitly,
\begin{equation}
P=p_0+p_1\zeta+\bar p_1\bar\zeta+p_2\zeta\bar\zeta\;,
\end{equation}
with ${p_0}$ and ${p_2}$ real constants, and ${p_1\in\mathbb{C}}$, satisfying (as a consequence of \eqref{Peqapx}) the relation
\begin{equation}\label{Pcoefcond}
p_0p_2-p_1\bar p_1=\frac12\Lambda_+\;.
\end{equation}
The solutions \eqref{P1} and \eqref{P2} are particular examples of such a choice.

Assuming \eqref{Plincond}, we can now easily find the reality condition for ${\lambda}$ given by \eqref{typeDlambdasol}. Since ${P}$ is real, it requires that ${Q\equiv P \lambda=(2\bar{\mathcal{L}}\,P_{,\bar\zeta}-\bar{\mathcal{L}}_{,\bar\zeta}\,P)/\Lambda_+}$ is also real. Taking the derivative ${Q_{,\zeta\zeta}}$, we find that it vanishes, and ${Q}$ is thus linear in both ${\zeta}$ and ${\bar\zeta}$:
\begin{equation}\label{Qlincond}
Q=q_0+q_1\zeta+\bar q_1\bar\zeta+q_2\zeta\bar\zeta\;.
\end{equation}
Here ${q_0}$ and ${q_2}$ must be real and ${q_1}$ complex transverse constants, but they can by ${u}$-dependent. The field equation \eqref{vaclambda} now gives the restriction
\begin{equation}\label{Qcoefcond}
p_0q_2+p_2q_0=p_1\bar q_1+\bar p_1 q_1\;.
\end{equation}
Substituting in \eqref{lambda,zeta}, we find
\begin{equation}
\begin{split}
\bar{\mathcal{L}}&=
  (p_0q_1-q_0p_1)+(\bar p_1 q_2-p_2\bar q_1)\bar\zeta^2\\
  &\quad+(p_0q_2-p_2q_0-p_1\bar q_1+\bar p_1 q_1)\bar\zeta\;.
\end{split}
\end{equation}

In particular, for ${P}$ given by \eqref{P1} 
we get
\begin{align}
P&=1+\frac12\Lambda_+\zeta\bar\zeta\;,\\
Q&=q_0(1-\frac12\Lambda_+\zeta\bar\zeta)+q_1\zeta+\bar q_1\bar\zeta\;,
\end{align}
which is equivalent to \eqref{gentypeDPQ} in real coordinates ${x}$ and ${y}$.
For ${P=\sqrt{-\Lambda_+}\,x}$, which is a generic example of the choice \eqref{P2}, we have ${Q=c\sqrt{-\Lambda_+}\,y}$ and
\begin{equation}
\lambda=-i c\,\frac{\zeta-\bar\zeta}{\zeta+\bar\zeta}=c\,\frac{y}{x}\;.
\end{equation}

Finally, we should solve the equations \eqref{vach} and \eqref{typeDh} for ${\hat h}$. As for ${\lambda}$, we find that ${\hat h_{,\zeta} = P^{-2}\bar{\mathcal{H}}}$, with ${\bar{\mathcal{H}}_{,\zeta}=0}$. By substituting into \eqref{vach}, this leads to
\begin{equation}
0=P^2\hat h_{,\zeta\bar\zeta}
 = \bar{\mathcal{H}}_{,\bar\zeta}-\bar{\mathcal{H}}\bigl(\log P^2\bigr)_{,\bar\zeta}\;.
\end{equation}
Taking the derivative with respect to ${\zeta}$ and using \eqref{Peqapx}, we find
\begin{equation}
0=\bar{\mathcal{H}}(\log P^2)_{,\zeta\bar\zeta}=\frac{\Lambda_+\bar{\mathcal{H}}}{P^2}\;.
\end{equation}
For ${\Lambda_+\neq0}$ we thus obtain that the conditions \eqref{vach} and \eqref{typeDh} admit only the trivial solution ${\bar{\mathcal{H}}=0}$. 
Thanks to \eqref{hath}, for ${h}$ we obtain
\begin{equation}
h=\frac12\Lambda_+\lambda^2-\frac12\Lambda_-\kappa^2-\partial_u\kappa\;,
\end{equation}
in which ${\kappa}$ can be set to zero by a proper gauge.

\subsection*{The case ${\Lambda_+=0}$}

If the transverse space is flat, i.e. ${\Lambda_+=0}$, we naturally chose Cartesian transverse coordinates for which ${P=1}$. We thus immediately get the conditions
\begin{equation}
  \lambda_{,\zeta\bar\zeta}=0\;,\quad   \lambda_{,\zeta\zeta}=0\;,
\end{equation}
which imply ${\lambda_{,\zeta}=q_1}$ and
\begin{equation}\label{lambdalin}
  \lambda=q_0+q_1\zeta+\bar q_1\bar\zeta=q_0+q_x x + q_y y\;,
\end{equation}
where ${q_0(u), q_x(u)}$ and  ${q_y(u)}$ are real and ${q_1(u)}$ complex functions of ${u}$ only. The metric 1-form ${a_i}$ is then
\begin{equation}
  a_i = (\kappa+q_y x - q_x y)_{,i }\;.
\end{equation}
Thanks to such a very special form \eqref{lambdalin}, it is a gradient, and therefore it could be transformed away by a suitable choice of ${\kappa}$ using the gauge transformation. However, it would generate a nontrivial function ${g}$ and contributions to the equation for ${\hat h}$. Therefore, such a gauge fixing may not be the best choice.

Since in this case ${\hat h}$ satisfies the same equation as ${\lambda}$, we can write
\begin{equation}\label{hathlin}
  \hat h=L_0+L_1\zeta+\bar L_1\bar\zeta=L_0+L_x x + L_y y\;,
\end{equation}
with ${L_0(u)}$, ${L_x(u)}$ and ${L_y(u)}$ real and ${L_1(u)}$ complex transverse constants. With the gauge ${\kappa=g=0}$, the relation \eqref{hath} gives
\begin{equation}\label{hlin}
  h=L_x x+L_y y\;,
\end{equation}
where the constant ${L_0}$ has been eliminated by rescaling the coordinate ${u}$, i.e., incorporating the gauge transformation \eqref{urep}. We have thus obtained the metric \eqref{genPlHac} which is the ${a_i\neq0}$ generalization of the exceptional Pleba\'nski--Hacyan spacetime.

\section{Scalar Green functions on constant curvature spaces}\label{apx:Green}

The Green functions for the Poisson and Helmholtz-Poisson equations \eqref{poisson} and \eqref{helmholtz} discussed in Section~ \ref{sc:GreenFc} are special cases of the Green function satisfying 
\begin{equation}\label{G1}
\left[\laplace-\xi R_\perp\right]\,G({x,x'})=-\delta({x,x'})\;,
\end{equation}
with ${\xi=0}$ and ${\xi=-1}$, respectively. In our case, the scalar curvature ${R_{\perp}=2\Lambda_+}$ is constant, and the manifold is either a 2-plane, a 2-sphere or a 2-hyperboloid, depending on the sign of $\Lambda_+$. Because of the symmetry of these spaces the Green function $G({x,x'})$ is only the function of the geodesic distance $\ell({x,x'})$ between the points ${x}$ and ${x'}$. In this appendix we present the Green functions for an arbitrary parameter~$\xi$. 

Since the case of flat plane is trivial, we will discuss only the cases of a hyperboloid and a sphere. Let us note that on a sphere, for particular values ${\xi=-\frac12 l(l+1)}$ with integer $l\ge 0$ the equation \eqref{G1} has to be modified in order to take into account normalizable zero modes. Namely, it is neccessary to subtract from the $\delta$-function a projector to the space of zero modes.

\subsection*{2-hyperboloid $H^2$}

For ${\Lambda_+<0}$, the Green functions can be expressed in terms of the Legendre functions $Q_{\nu}$ (see, e.g., \cite{Zelnikov:2008:JHEP:})
\begin{equation}\label{G2}
G_{(\nu)}({x,x'})={\frac1{2\pi}}\,Q_{\nu}(\eta)\;,
\end{equation}
where
\begin{gather}
\nu=-{\frac12}+\sqrt{{\frac14}-2\xi}\;,\label{G3a}
\end{gather}
and ${\eta(x,x')}$ is given by the equations \eqref{h2}, \eqref{h3}.
When the separation between the two points increases, $\eta\rightarrow\infty$ and the Green function \eqref{G2} tends to zero. This is a proper boundary condition for Green functions on hyperbolic spaces.
In our paper only the ${\nu=0,1}$ cases are important, and the corresponding Green functions can be expressed in terms of elementary functions:
\begin{equation}\label{G5}
G_{(0)}({x,x'})={\frac1{2\pi}}Q_{0}(\eta)=-{\frac1{4\pi}}\log\Bigr(\frac{\eta-1}{\eta+1}\Bigr)\;,
\end{equation}
\begin{equation}\label{G6}
G_{(1)}({x,x'})={\frac1{2\pi}}Q_{1}(\eta)=-{\frac1{4\pi}}\left(\eta\log\Bigr(\frac{\eta-1}{\eta+1}\Bigr)+2\right)\;.
\end{equation}

\subsection*{2-sphere $S^2$}

When ${\Lambda_+>0}$, the Green function \eqref{G1} for
arbitrary $\xi$ can be expressed in terms of the Legendre function ${\rm P}_{\nu}$:
\begin{equation}\begin{aligned}\label{G7}
G_{(\nu)}({ x,x'})&=-{\frac1{4\sin{\pi\nu}}}{\rm P}_{\nu}(-\eta)\\
&={\frac1{2\pi}}{\rm Q}_{\nu}(\eta)-{\frac14}\cot({\pi\nu})\,{\rm P}_{\nu}(\eta)\;,
\end{aligned}\end{equation}
see \cite{Zelnikov:2008:JHEP:}, with ${\nu}$ given by \eqref{G3a} and ${\eta(x,x')}$ by the equations \eqref{s3} and \eqref{s4}.
The Legendre functions ${\rm P}_\nu,\,{\rm Q}_\nu$ are defined on the interval $[-1,1]$ and are related to the Legendre functions ${P_\nu,\,Q_\nu}$ on the complex plane as ${{\rm P}_\nu(\eta)=\frac12[P_\nu(\eta{+}i0)+P_\nu(\eta{-}i0)]}$, and similarly for ${Q}$s \cite{Bateman:book}. When ${0<\eta<1}$ we can observe from the second line of \eqref{G7} that there is the required singularity at ${\eta=1}$, when both points coincide. For ${-1<\eta<0}$, the first line gives the regularity of the Green function at ${\eta=-1}$ corresponding to antipodal points.


If $\nu$ is an integer number, one has to be cautious because of a contribution of normalizable zero modes. The zero modes must be removed from the spectrum. This case thus must be treated separately.

When $\nu=0$, the corresponding Green function then satisfies the equation
\begin{equation}\begin{aligned}\label{G10}
\laplace G_{(0)}({x,x'})=-\delta({x,x'})+{\frac{\Lambda_+}{4\pi}}\;,
\end{aligned}\end{equation}
where ${\frac{4\pi}{\Lambda_+}}=\int_{S^2}\sqrt{g_{\perp}}{d^2{x}}$ is the total area of the compact transverse space \eqref{trmetric}. 
The term subtracted from the delta function is exactly the projector on the zero mode, which for ${\nu=0}$ is constant.

The solution of this problem, which is regular at antipodal points ($\eta=-1$) reads
\begin{equation}\begin{aligned}\label{G11}
G_{(0)}({x,x'})&=-{\frac1{4\pi}}\log\left(1-\eta\right)\;,
\end{aligned}\end{equation}
with $\eta$ given again by \eqref{s3}, \eqref{s4}. One can shift this Green function by a constant to satisfy different boundary conditions at the antipodal points, but this constant is unimportant since it will drop out of the equation \eqref{heq} anyway due to the constraint \eqref{s6}.

In the case of the equation \eqref{rotjeq} we have $\nu=1$. 
In this case the projector on zero modes is proportional to ${\rm P}_1(\eta({ x,x'}))=\eta({x,x'})$ and the equation for the Green function reads
\begin{equation}\begin{aligned}\label{G12}
\left[\laplace +2\Lambda_+\right]\,G_{(1)}({x,x'})=-\delta({x,x'})+\frac{3\Lambda_+}{4\pi}\,\eta({x,x'})\;.
\end{aligned}\end{equation}
The solution can be found explicitly as
\begin{equation}\begin{aligned}\label{G13}
G_{(1)}({x,x'})&=-{\frac1{4\pi}}\Bigl(\eta\log\left({1-\eta}\right)+1\Bigr)\;.
\end{aligned}\end{equation}
Similarly to the ${\nu=0}$ case, one can add to the Green function the projector to the zero modes with an arbitrary coefficient to match different boundary conditions at the antipodal point. But the component $T^{\rm gyr}_{ui}=j_{i}=\epsilon_{i}{}^k q_{,k}$ of the stress-energy tensor of the gyraton has to satisfy the integral constraint similar to \eqref{s6}, namely
\begin{equation}\begin{aligned}\label{G14}
\int_{S^2} \cos\theta\;q\,\sqrt{g_{\perp}}\, {d^2{x}}=0\;.
\end{aligned}\end{equation}
Because of this property, zero modes do not contribute to the components $a_i$ of the metric.
Constraints \eqref{s6}, \eqref{s7} appear only because $S^2$ is a compact manifold.

Let us finally note that in the case of a general integer ${\nu=l}$ the modified Green function equation is
\begin{equation}\label{G15}
\begin{split}
&\left[\laplace\!\!+l(l{+}1)\Lambda_+\right]\,G_{(l)}({ x,x'})\\
&\qquad=-\delta({ x,x'})+\Lambda_+\frac{2l{+}1}{4\pi}\,{\rm P}_l(\eta(x,x'))\;.
\end{split}
\end{equation}
Its solution can be written as
\begin{equation}\label{G16}
G_{(l)}({x,x'})=-{\frac1{4\pi}}\Bigl({\rm P}_l(\eta)\log\left({1-\eta}\right)+U_l(\eta)\Bigr)\;,
\end{equation}
with $U_l(\eta)$ being a specific polynomial in $\eta$ of the \mbox{${(l{-}1)}$-th} order. 
In particular, 
${U_0=0}$,
${U_1=1}$,
${U_2=\frac{3}{2}\eta+\frac{1}{3}}$,
${U_3=\frac{5}{2}\eta^2+\frac{3}{5}\eta-\frac{2}{3}}$, etc.




\end{document}